%% file: main.tex
\documentclass{article}

\usepackage{soul}
\usepackage{amsmath} 
\usepackage{arxiv}
\usepackage{xcolor}
\usepackage[textsize=footnotesize]{todonotes}
\usepackage[font=small,skip=0pt]{caption}

\input{includes}

\title{Exploring the Advantages of Transformers for High-Frequency Trading}

\author{ Fazl Barez \\
        Edinburgh Centre for Robotics \\
        The University of Edinburgh\\
        \texttt{f.barez@ed.ac.uk}\\
    \And
    Paul Bilokon \\
	Department of Computing \\
	Imperial College London \\
	\texttt{paul.bilokon@imperial.ac.uk} \\
	\And
	Arthur Gervais \\
	Department of Computing \\
	Imperial College London \\
	\texttt{a.gervais@imperial.ac.uk}
        \And
    	\And
	Nikita Lisitsyn \\
	Department of Computing \\
	Imperial College London \\
	\texttt{nikita.lisitsyn21@imperial.ac.uk}\\}

\input{notation}

\usepackage[skip=5pt plus1pt, indent=20pt]{parskip}
\usepackage[numbers]{natbib}
\setcitestyle{square}
\usepackage{multirow, makecell}
\usepackage{booktabs}
\usepackage{setspace}
\usepackage{titlesec}
\usepackage{fancyhdr}
\usepackage{float}

\begin{document}

\maketitle

\begin{abstract}
This paper explores the novel deep learning Transformers architectures for high-frequency Bitcoin-USDT log-return forecasting and compares them to the traditional Long Short-Term Memory models. A hybrid Transformer model, called \textbf{HFformer}, is then introduced for time series forecasting which incorporates a Transformer encoder, linear decoder, spiking activations, and quantile loss function, and does not use position encoding. Furthermore, possible high-frequency trading strategies for use with the HFformer model are discussed, including trade sizing, trading signal aggregation, and minimal  trading threshold. Ultimately, the performance of the HFformer and Long Short-Term Memory models are assessed and results indicate that the HFformer achieves a higher cumulative PnL than the LSTM when trading with multiple signals during backtesting\footnote{The code will be publicly available upon publication}.
\end{abstract}

\section{Introduction}

Forecasting Financial Time Series (FTS) has been of interest to financial market participants who are interested in making profitable trades on the financial markets. It has historically been approached using stochastic and machine learning models. Stochastic methods include linear models such as Autoregressive Integrated Moving Average (ARIMA)~\cite{box2015time} that support non-stationary time series and non-linear models, including the Generalized Autoregressive Conditional Heteroskedasticity (GARCH)~\cite{engle1982autoregressive} model. Machine learning methods are data-driven approaches, among which Recurrent Neural Networks (RNNs) ~\cite{rumelhart1985learning}, more specifically, Long Short-Term Memory (LSTM) networks ~\cite{hochreiter1997long}, have been especially popular for time series prediction. Periodically, new deep learning models are being adopted in quantitative research to find the most accurate models in FTS forecasting that would lead to more efficient trading strategies. 

Recently, a new type of deep learning~\cite{lecun2015deep} architecture called Transformer~\cite{vaswani2017attention}, relying on Attention~\cite{bahdanau2014neural}, was introduced for Natural Language Processing (NLP) applications. Transformers have since been used in other applications such as computer vision tasks~\cite{liu2021swin} and more recently in time series forecasting. This paper will focus on the application of Transformers in high-frequency FTS forecasting. 

FTS are characterized by properties including frequency, auto-correlation, heteroskedasticity, drift, and seasonality \cite{tsay2005analysis}. The frequency of trading strategies ranges from milliseconds to minutes and days. Most studies on FTS forecasting~\cite{kolm2021deep, qin2017dual} focus on minute frequencies as millisecond trading data is expensive. For the results of this paper to be in line with real-world trading conditions, we will use millisecond frequency tick price and level 2 Limit Order Book (LOB) data~\cite{cartea2015algorithmic} which will allow us to focus on the bid and ask prices and quantities of the first ten levels of the LOB. We will be using data collected from the  Binance cryptocurrency exchange for the Bitcoin - USDT trading pair. 

Traders require automated systems to make decisions based on the order book and tick price data at millisecond frequencies. Stochastic modeling has long been used to describe prices as exogenous random variables, however, it relies on assumptions that may oversimplify the complexity of financial markets.  Contrary to stochastic modeling, deep learning forecasting relies on a data-driven approach to learn past dynamics to predict future behavior. Deep learning approaches have shown promising results in High-Frequency Trading (HFT) strategies~\cite{kolm2021deep}. Currently, LSTMs are the most widely used deep learning architecture for FTS forecasting \cite{sezer2020financial}. However, LSTMs are affected by vanishing and exploding gradient problems for long time series~\cite{pascanu2013difficulty}, and they are sequential structures that are difficult to parallelize. 

Transformers use multi-head Attention to capture long-term dependencies and are naturally parallelizable, which makes them a good candidate for low-latency trading strategies. However, Transformers were not specifically engineered for modeling local sequential structures and require finely tuned position embeddings. Modeling short- and long-range temporal dependencies while accounting for FTS properties (e.g., seasonality and auto-correlation) remains an open question~\cite{xu2021autoformer,zhou2022fedformer}.   

In this paper, we compare the performance of existing deep learning architectures for FTS forecasting. We then introduce an improved deep learning model called the HFformer, which combines certain features of previously compared architectures. Finally, we compare the performance of the HFformer to existing architectures in the context of FTS forecasting. We focus on describing the collected cryptocurrency data and detailing the pre-processing of the data before feeding it into deep learning models. Then, we cover the LSTM and original Transformer encoder-decoder architecture and compare their forecasting performance. We focus on understanding more recent Transformer architectures precisely engineered for time series forecasting: Autoformer and FEDformer. We evaluate the idiosyncrasies of these novel architectures and compare their performance to the original Transformer. We combine some novelties introduced by time series Transformer architectures to create a Transformer specifically optimized for HFT, which we refer to as HFformer. Finally, we compare the HFformer's performance to the most currently used deep learning architecture for FTS forecasting, the LSTM, on larger datasets. We also compare the trades done by both architectures. We conclude by presenting possible trading strategies and assessing their performance. 

\section{Background and Related Work}

\subsection{FTS}

The world financial markets have undergone a technological revolution in the past decades. The execution time of market orders decreased from 25 ms in 2000 to less than 0.017 ms in 2017~\cite{angel2018retail}. The trades are spaced randomly and endogenously~\cite{hautsch2011modelling} as more trading occurs during market openings and closings, breaking news, and abrupt price changes. This leads to auto-correlated prices and quantities. FTS are non-stationary as they exhibit seasonality at different scales linked to the market opening hours, underlying assets, and economic events. Properties such as trend, drift, auto-correlation, and heteroskedasticity make FTS forecasting a challenging problem. 

\subsection{Traditional Methods for FTS}

Stochastic FTS forecasting relies on linear and non-linear methods. Linear models include Moving Average (MA), Autoregressive (AR), and ARIMA \cite{ariyo2014stock} models. Linear models can solve stationary and non-stationary time series. Non-linear models such as GARCH and GARCH-based models are used to forecast stock market volatility exhibiting heteroskedasticity \cite{franses1996forecasting}.

Currently, LSTMs are successfully used for minute-frequency FTS forecasting ~\cite{fischer2018deep} for one- and multi-step forecasting~\cite{lim2021time}. Convolutional Neural Networks (CNNs) and LSTM-CNNs are used to extract information from LOBs~\cite{kolm2021deep}. Spiking Neural Networks (SNNs) predict price spikes for HFT strategies~\cite{gao2021high}.  

Attention layers have been combined with LSTMs~\cite{wang2016attention} for NLP sentiment classification tasks. Recently, LSTM encoder-decoder architectures with dual Attention have been used to make the encoder and decoder hidden state selection more relevant for FTS forecasting~\cite{qin2017dual}. 

\subsection{Transformers for FTS}

Initially introduced for NLP, the Transformer has shown state-of-the-art performance for forecasting time series~\cite{wu2020deep}. The Transformer requires a positional encoding, such as Time2Vec \cite{kazemi2019time2vec}, to be added to the input embeddings to model time series.  Novel Transformer designs~\cite{li2019enhancing} have focused on reducing the $O(L^2)$ complexity by using causal convolutions or sparse bias in the Attention module to process longer sequences.

Other structural changes were made to improve time series forecasting. The Autoformer~\cite{xu2021autoformer} has a modified auto-correlation Attention module that selects the top-$k$ similar sub-sequences. The Temporal Fusion Transformer (TFT)~\cite{lim2019temporal} uses static covariate encoders, gating layers for noise suppression, and  recurrent layers for local processing. The FEDformer~\cite{zhou2022fedformer} performs a Fourier and Wavelet time series decomposition using the Attention mechanism.

\section{Exploratory Data Analysis}

\subsection{FTS data}
Level 2 LOB data is collected from the Binance cryptocurrency exchange for the BTC-USDT trading pair using Binance's 100 ms LOB web socket\footnote{More details on Binance's LOB web socket can be found at \href{https://www.binance.com/en/binance-api}{https://www.binance.com/en/binance-api}.}.

Most of the collected LOB snapshots for the BTC-USDT trading pair are within 100 ms. We then proceed to compute the weighted midprice defined as follows:
\begin{gather}
Midprice_{t} = \frac{q_{t}^{a;1} \cdot p_{t}^{a;1} + q_{t}^{b;1} \cdot p_{t}^{b;1}}{2}
\end{gather}
where $p_{t}^{a;1}$, $q_{t}^{a;1}$, $p_{t}^{b;1}$, and $q_{t}^{b;1}$ are the bid price and quantity and ask price and quantity of level 1 of the LOB at time $t$.

We process the collected LOB data by dropping snapshots with the same consecutive midprice value as the analysis will be done at trade time rather than wall clock time. This is done because we are interested in predicting the values at which the future trades will occur, and this brings the distribution of the observed values closer to a normal distribution (cf. \autoref{sec_logreturns}). In this paper, the terms ``midprice" and ``weighted midprice" will be used interchangeably. 

\subsubsection{Log-Returns}\label{sec_logreturns}

One challenge with FTS is that the range of values varies significantly, even for the same financial asset. If we want to apply a deep learning method to forecast FTS, we need to transform the time series to remove the non-stationarity. Usually, three price time series transformations are proposed:
\begin{itemize}
    \item price differencing: $d_{t+1} = p_{t+1} - p_{t}$ 
    \item returns: $r_{t+1}^{*} = \frac{p_{t+1} - p_{t}}{p_t}$ 
    \item log-returns: $r_{t+1} = \log(p_{t+1}) - log(p_{t})$
\end{itemize}

All three of these methods remove the non-stationarity of prices. However, using price differencing preserves the units and will be proportional to the underlying price of the asset, which is undesirable in the case of Bitcoin because of its high volatility. On the contrary, returns and log-returns are unitless as it is a ratio of two prices. Returns represent the change in asset price as a percentage and are therefore widely used in finance for their convenience.

From now on, we focus on forecasting log-returns over different time horizons. We define log-returns the following way:
\begin{gather}
r_{t+\tau} = \log(p_{t+\tau})-\log(p_{t}) = \log\left( \frac{p_{t+\tau}}{p_{t}}\right)
\end{gather}
where $\tau \in \{1,...,30\}$.
$\tau$ is the value of the forecasting horizon.

\begin{figure}[H]
\centering 
\includegraphics[width=0.8\linewidth]{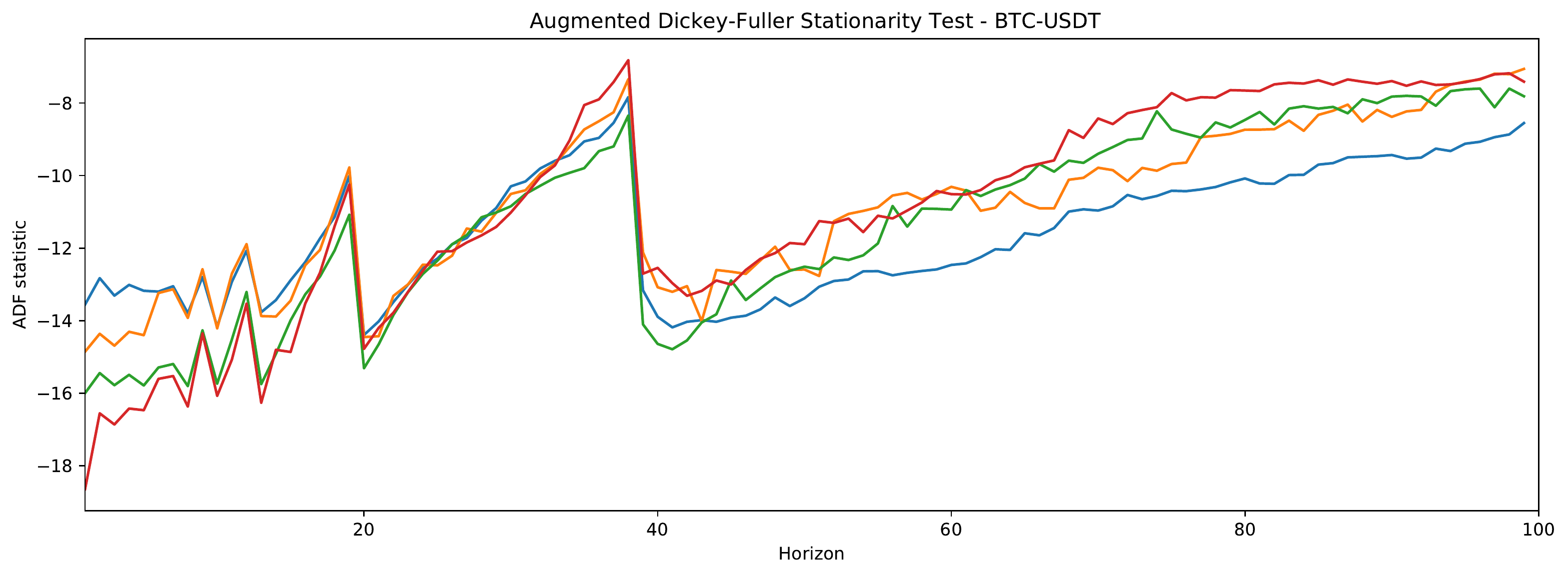} 
\caption{Augmented Dickey-Fuller stationarity test for different log-return horizons of four different samples of 9000 LOB snapshots.}
\label{fig2_1}
\end{figure}

The range of considered forecasting horizons can be justified using the Augmented Dickey-Fuller (ADF) stationarity test, which tests for different horizons of the log-returns. As the prediction horizon increases, the information contained in the observed prices, quantities, and historical returns becomes less informative for future predictions (cf. \autoref{fig2_1}) because the time series becomes less stationary. A lower value means that the time series exhibits stronger stationary properties. 

\subsubsection{LOB}

\begin{figure}[ht]
\centering 
\includegraphics[width=0.8\linewidth]{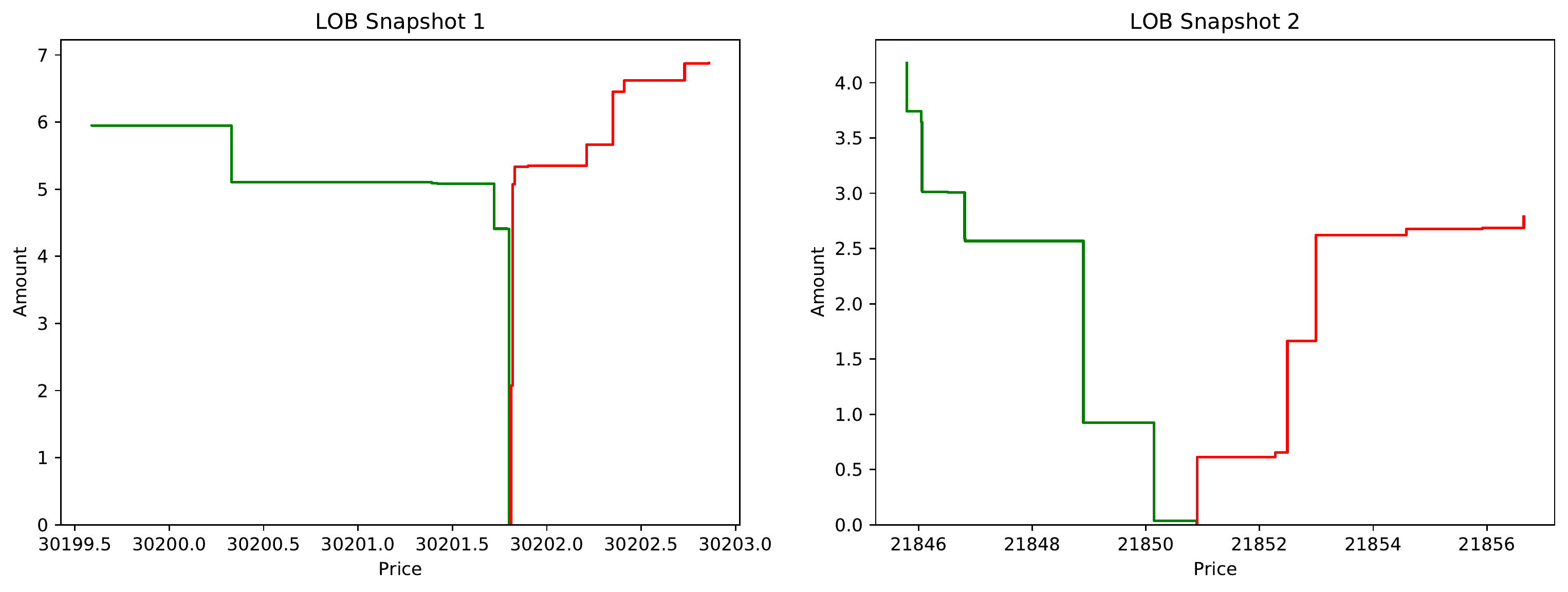} 
\caption{Bid-ask pressure for 10 levels of bid and ask for BTC-USDT.}
\label{fig3_1}
\end{figure}

The LOB contains information about the bid and ask prices and the quantities offered at those prices (cf. \autoref{fig3_1}). Market participants can choose to place different types of orders. Market orders are executed at the current market price if the offered quantity allows; otherwise, the order will be partially filled until completion at higher levels of the LOB for buy orders and lower levels of the LOB for sell orders. This can sometimes lead to price slippage, resulting in a less desirable average fill price. Another strategy is to place a limit order which will be executed only at the specified price or better as mentioned in the order. Depending on the limit price, these orders will appear at a certain level in the LOB.

\autoref{fig3_1} shows two examples of the bid-ask pressure. In the first case, the midprice will have a greater resistance to change than in the second case since a larger amount of BTC would need to be bought or sold to move the midprice.

\begin{figure}[ht]
\centering 
\includegraphics[width=0.8\linewidth]{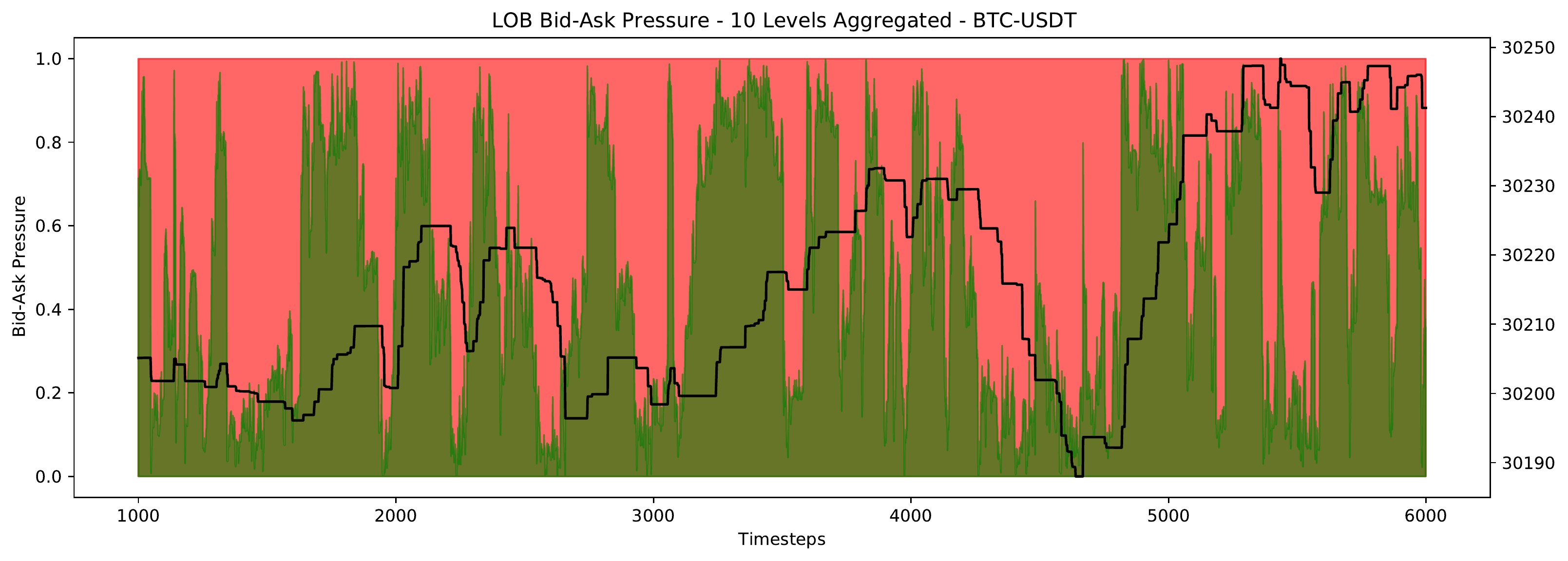} 
\caption{LOB bid-ask pressure for 10 aggregated levels.}
\label{fig3_2}
\end{figure}

From \autoref{fig3_2}, we notice that as the bid pressure builds up in the LOB, the weighted midprice starts to move higher and vice-versa for the ask pressure. This leads us to explore the predictive power of the information contained in the LOB for forecasting future log-returns. 

\subsubsection{Normalization of FTS}

To improve the performance of machine learning methods, input data needs to be normalized. Traditionally, the normalization parameters from the training set are applied to the test set (e.g., the mean and standard deviation of the training data). However, as in the case of the BTC-USDT pair, the mean and standard deviation of the prices and quantities fluctuate significantly. Taking this into account, we propose to use online normalization. 

The online normalization of the input data is done by computing the mean and standard deviation of the data contained in the look-back window. We then normalize the data in the look-back window using the computed mean and standard deviation. Later, we will test the performance of the forecasting models for look-back windows of sizes 100, 200, 300, and 400 timesteps (cf. \autoref{fig6_1_2}). During data normalization, it is important not to compute any estimates based on future values that we want to forecast to avoid contaminating the training set.   

\section{LSTM and Transformers for FTS}

\subsection{LSTM}

Recurrent Neural Networks (RNNs) are neural networks that were developed to work with sequential data to model dependencies of arbitrary length. The goal is to learn the mapping between $x_{1:T}$ and $y_{1:T}$ by encoding the input sequence into a hidden state $h_{t}$ that is then decoded into $y_{t}$:
\begin{gather}
\vec{h}_{t} = \phi_{h}(W_h\vec{h}_{t-1}+W_x\vec{x_t}+\vec{b}_h) \\
\vec{y}_{t} = \phi_{y}(W_y\vec{h}_{t}+\vec{b}_y)
\end{gather}
where $W_h, W_x, W_y, \vec{b}_h, \vec{b}_y $ are learned parameters and $\phi_{h}$ and $\phi_{y}$ are activation functions.

However, RNNs suffer from vanishing/exploding gradients due to the backpropagation mechanism used during Gradient Descent (GD). GD is an iterative optimization algorithm used to find a given function's local optimum to reduce the difference between the predicted and true values. Various loss functions are used to compute the difference between the true and predicted values, which is then propagated back to modify the weights of the model. This issue is mitigated by using memory cell states and gating mechanisms in LSTMs:

\begin{gather}
\vec{f}_{t} = \sigma(W_f \cdot [\vec{h}_{t-1}, \vec{x_t}]+\vec{b}_f) \\
\vec{i}_{t} = \sigma(W_i \cdot [\vec{h}_{t-1}, \vec{x_t}]+\vec{b}_i) \\
\vec{o}_{t} = \sigma(W_o \cdot [\vec{h}_{t-1}, \vec{x_t}]+\vec{b}_o) \\
\vec{\Tilde{c}}_{t} = \tanh(W_c \cdot [\vec{h}_{t-1}, \vec{x_t}]+\vec{b}_c) \\
\vec{c}_{t} = \vec{f} \odot \vec{c}_{t-1} + \vec{i}_{t} \odot \vec{\Tilde{c}}_{t} \\
\vec{h_t} = \vec{o_t} \odot \vec{c_t}
\end{gather}
where $W_f, W_i, W_o, W_c, b_f, b_i, b_o, b_c$ and $\sigma{\cdot}$ is the sigmoid activation function.
\subsection{Transformer}

\subsubsection{Attention}
A scaled dot-product Attention~\cite{vaswani2017attention} mechanism is a key-value lookup based on a query matrix $\vec{Q}$, key matrix $\vec{K}$, and value matrix $\vec{V}$ described as:
\begin{gather}
Attention(\vec{Q},\vec{K},\vec{V};\alpha) = \alpha\left(\frac{\vec{Q}\vec{K}^T}{\sqrt{d_q}}\right)\vec{V} 
\end{gather}
where $\vec{Q} \in \R^{N \times d_q}, \vec{K} \in \R^{M \times d_q}, \vec{V} \in \R^{M \times d_v},$ and $\alpha(.)$ is an activation function applied row-wise. $d_q$ and $d_v$ represent the dimensions of the query and value vectors respectively. The $softmax(.)$ activation function is used in Transformers and is defined as:
\begin{gather}
softmax(x_{i}) = \frac{\exp(xi)}{\sum_{j}\exp(x_j)}
\end{gather}
Intuitively $softmax(.)$ transforms a vector $\vec{x}$ into a probability vector for which the sum of elements adds up to 1.

Transformers, compared to RNNs, are easily parallelizable and can be fed multiple consecutive time series in a batch to encode and decode. However, providing consecutive time series may prevent the Transformer from learning, as it could use the time series in the batch to see the ``future." Masked Attention prevents the Transformer from seeing future values:
\begin{gather}
MaskedAttention(\vec{Q},\vec{K},\vec{V};\alpha, \vec{M}) = \alpha\left(mask\left(\frac{\vec{Q}\vec{K}^T}{\sqrt{d_q}}, \vec{M}\right)\right)\vec{V}
\end{gather}
where $\vec{M}_{nm}$ is $0$ for masking and $1$ otherwise. When $\vec{M}_{nm} = 0$, set $(\vec{Q}\vec{K}^T)_{nm} = -\infty$ which in the case of $softmax(.)$ will result in $\exp\left(\frac{\vec{Q}\vec{K}^T}{\sqrt{d_q}}\right) = 0$. Therefore, the value $\vec{v}_{m}$ will not contribute to the Attention output for query $\vec{q}_{n}$.

Transformers combine multiple Attention heads into a multi-head Attention module where
each head captures different parts of the sequence, which allows for coarser and finer encodings of the time series.

\subsubsection{Position Encoding}
Notice that scaled dot-product Attention is permutation equivariant:
\begin{gather}
Attention(\vec{P}\vec{Q},\vec{K},\vec{V};\alpha) = \alpha\left(\frac{\vec{P}\vec{Q}\vec{K}^T}{\sqrt{d_q}}\right)\vec{V} = \vec{P}\alpha\left(\frac{\vec{Q}\vec{K}^T}{\sqrt{d_q}}\right)\vec{V}
= \vec{P}Attention(\vec{Q},\vec{K},\vec{V};\alpha)
\end{gather}
As the ordering information matters in time series forecasting, where we input a sequence of queries $\vec{q}_1, \dots, \vec{q}_N$, we construct a modified version of the query matrix $\vec{Q}$:
\begin{gather}
\vec{\widetilde{Q}} = (\vec{\widetilde{q}}_1, \dots, \vec{\widetilde{q}}_N)^T,\quad \vec{\widetilde{q}}_n = f(\vec{q}_n, PE(n))
\end{gather}
where $f(.)$ is a summation and concatenation, and $PE(n)$ is the position encoding of input with index $n$, which could be learned or computed. We use sinusoid embedding to encode FTS ordering information. 

\subsubsection{Layer Normalization}
Layer normalization~\cite{ba2016layer} is used to stabilize the training of the neural network. 

Let $\vec{y} = \{y_1,\dots,y_n\}= \vec{W}\vec{x}+\vec{b}$ be the output of a feed-forward layer before going through an activation function. The layer normalization is defined as:
\begin{gather}
    y \leftarrow \frac{y-\hat{y}}{\sigma}, \quad \hat{y} = \frac{1}{n}\sum_{k=1}^{n}y_{k}, \quad \sigma=\sqrt{\frac{1}{n}\sum_{k=1}^{n}(y_{k}-\hat{y})^2}
\end{gather}
In Transformers, layer normalization is applied with a residual connection: 
\begin{gather}
Add\&Norm(x) = LayerNorm(x + Sublayer(x))
\end{gather}
where $Sublayer(.)$ can either be a multi-head Attention layer or a point-wise feed-forward network.

\subsubsection{Point-wise Feed-forward Network}
Transformers use the point-wise feed-forward network as the feed-forward layer. The output of the multi-head Attention layer after the Add\&Norm layer is a $N \times d_{out}$ matrix which contains the Attention results for $N$ queries. The point-wise feed-forward network processes this matrix row by row.

\subsection{Novel Transformer Architectures}

\subsubsection{Autoformer}
The Autoformer uses a new type of Attention based on auto-correlation in its encoder and decoder. The encoder eliminates the long-term trend by series decomposition using the Fast Fourier Transform and focuses on modeling seasonal patterns. The past seasonal and trend information from the encoder is then used by the decoder to predict future values.

\begin{figure}[H]
  \centering
  \begin{minipage}[b]{0.495\hsize}
    \includegraphics[width=\textwidth]{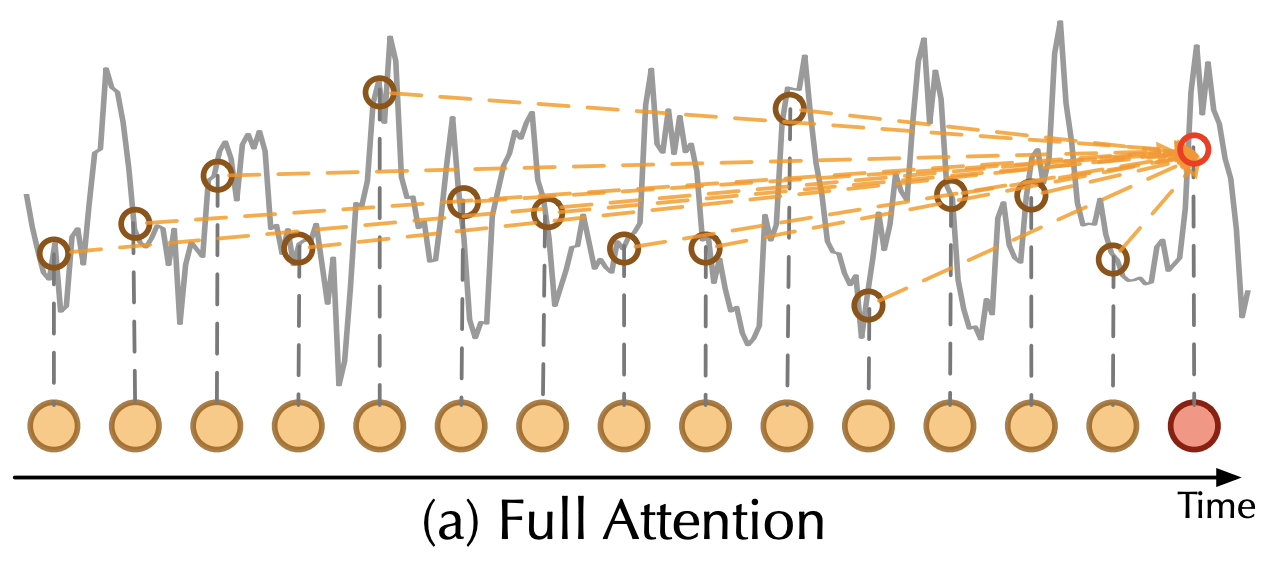}
    \caption{Attention module in a Transformer \cite{xu2021autoformer}.}
    \label{4_1_1_1}
  \end{minipage}
  \hfill
  \begin{minipage}[b]{0.495\hsize}
    \includegraphics[width=\textwidth]{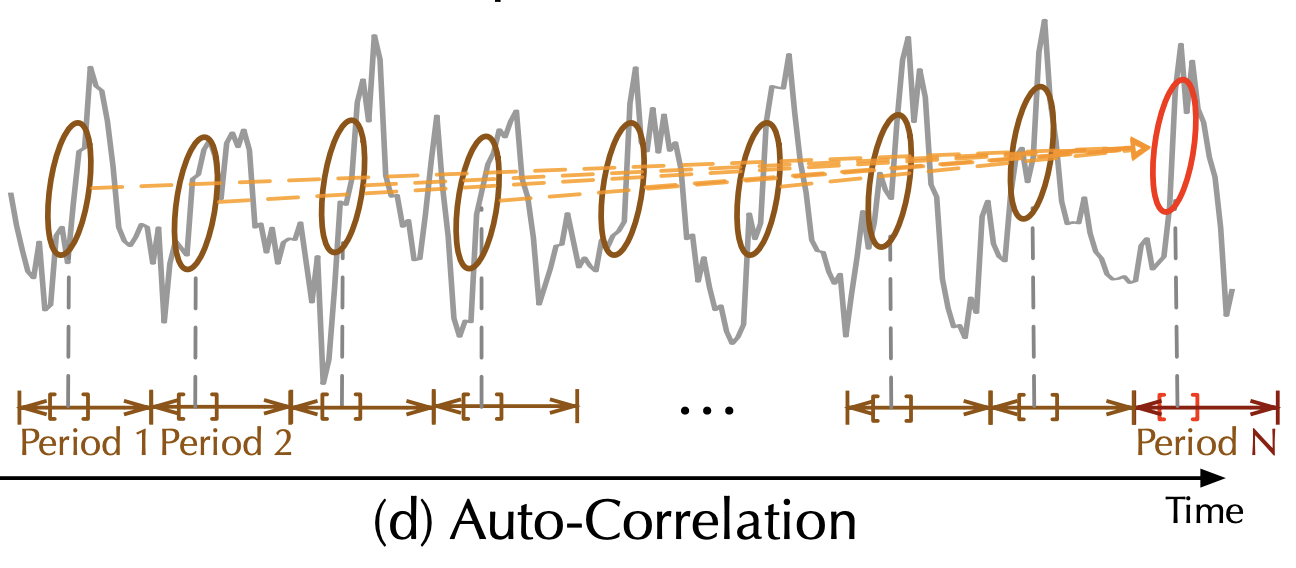}
    \caption{Auto-correlation Attention module in Autoformer \cite{xu2021autoformer}.}
    \label{4_1_1_2}
  \end{minipage}
\end{figure}

Compared to the Transformer's full Attention module (cf. \autoref{4_1_1_1}), the auto-correlation Attention (cf. \autoref{4_1_1_2}) achieves a reduced complexity of $O(L\log(L))$ compared to $O(L^2)$ for a length-$L$ time series. The performance improvement is due to querying only the top $\log(L)$ entries based on the series periodicity.

\subsubsection{FEDformer}

The FEDformer, similar to the Autoformer, uses seasonal-trend decomposition to predict future values. The FEDformer uses either the Fast Fourier Transform or the Wavelet Transform to map the time series into the frequency domain. It then randomly queries the frequency domain Attention module for high-frequency and low-frequency components, which reduces the complexity from $O(L^2)$ to $O(L)$ for length-$L$ time series. The goal behind using the frequency domain is to achieve an optimal mix between high-frequency components, which are more affected by local noise, and low-frequency components, which are more affected by the overall trend of the time series. 

\subsection{HFformer}

Having evaluated existing Transformer-like architectures, we combine some of their different architectural components into a new Transformer-like architecture called the HFformer. Below is an illustration of the architecture of the HFformer.

\begin{figure}[ht]
\centering 
\includegraphics[width=0.8\linewidth]{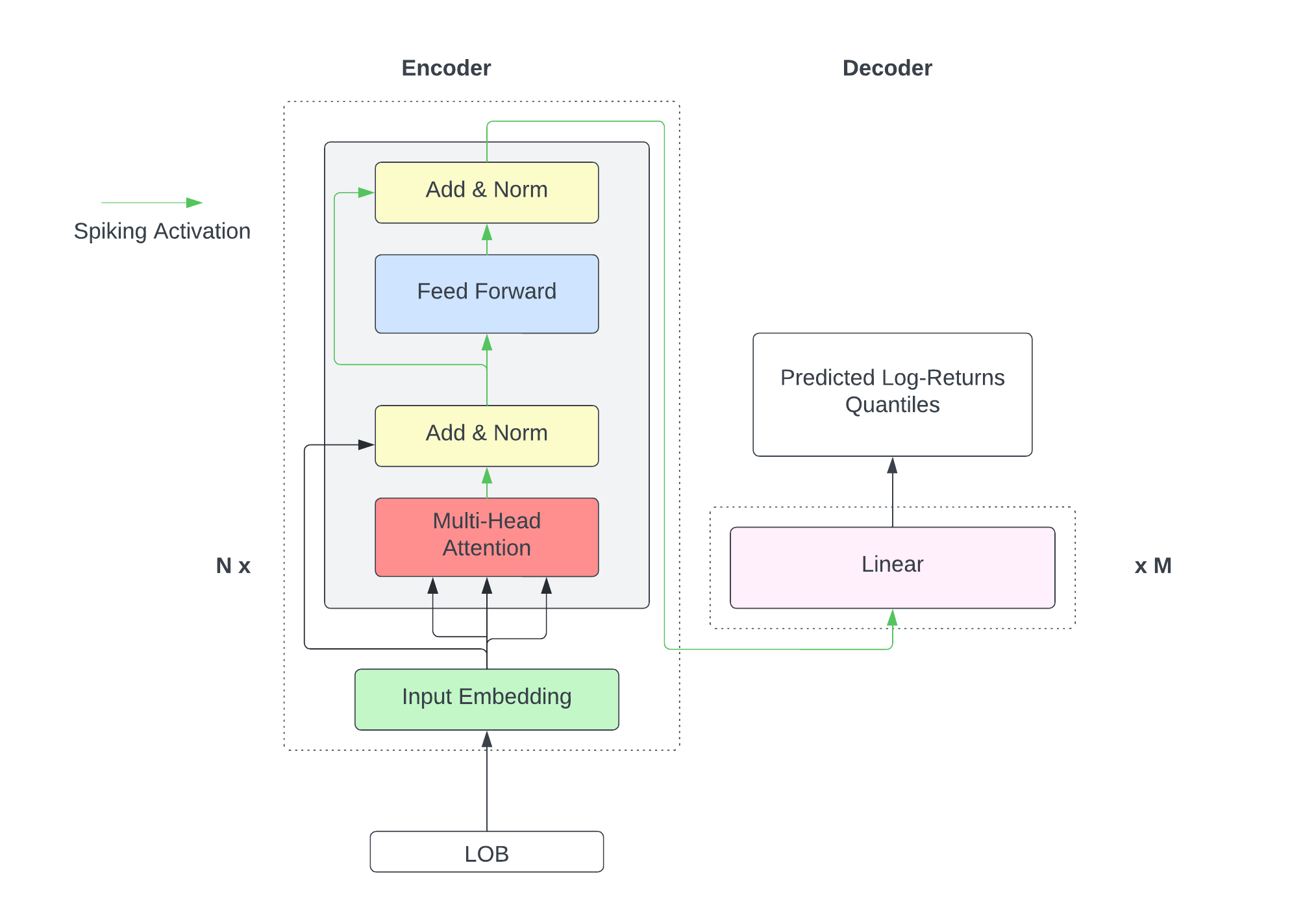} 
\caption{HFformer - Architecture.}
\label{fig5_1_1}
\end{figure}

\subsubsection{Transformer Encoder with Spiking Activation}
The HFformer's encoder is based on a Transformer encoder with activation functions that were modified to be spiking:
\begin{gather}
\textit{SpikingActivation}(x) =  
\begin{cases}
            x, &         \text{if } x \ge \textit{threshold},\\
            0, &         \text{if } x < \textit{threshold}.
    \end{cases}
\end{gather}
where the $threshold$ is learned during training. 

We use a combination of the spiking activation with a PReLU activation function. The Pytorch-Spiking Python library\footnote{More details about the Pytorch-Spiking Python library can be found at \href{https://github.com/nengo/pytorch-spiking}{https://github.com/nengo/pytorch-spiking}.} provides the implementation of the spiking activation. Due to the noisy nature of the LOB data, a spiking activation function reduces the noise propagated through the Transformer encoder by learning a passing threshold below which the information is not propagated in the encoder. 

\subsubsection{Linear Decoder}
The linear decoder contains two linear layers connected by a PReLU activation function. The decoder needs to output a single value for each forecast horizon as we forecast log-returns. Shifting from an autoregressive decoder to a linear decoder helps to reduce error propagation and decrease the training time. 

\subsubsection{No Positional Encoding}
The positional encoding was removed from the model. As a result, the model's performance was improved, and the training time was reduced as the number of trainable parameters decreased. Positional encoding was introduced for NLP tasks. It is used as an augmentation technique for the self-Attention mechanism, which is invariant to the sequence order. However, as we no longer require an autoregressive decoder since we use a linear decoder, the positional encoding becomes less relevant. Additionally, the amount of information provided to the Transformer encoder increases with the position dimension, which could serve as a form of positional encoding ~\cite{irie2019language}.

\subsubsection{Quantile Loss}
The quantile loss is defined as follows:
\begin{gather}
L(\hat{y}_{i}^{p}, y_i) = max[q(\hat{y}_i^p - y_i), (q-1)(\hat{y}_i^p - y_i)]
\end{gather}
where $\hat{y}_{i}^{p}$ is the model's output for a given quantile $p$ and $y_i$ is the true value. The mean reduction is used for the loss.

The HFformer works with MSE and MAE loss. However, quantile loss allows forecasting an interval around the predicted value. This is helpful when the inter-quartile range becomes larger than the mean inter-quartile range, which could signify a low certainty about the forecasted value and therefore one might decide not to enter the trade. 

\section{Experiments}

\subsection{Data}
\begin{figure}[H]
  \centering
  \begin{minipage}[b]{0.495\hsize}
    \includegraphics[width=\textwidth]{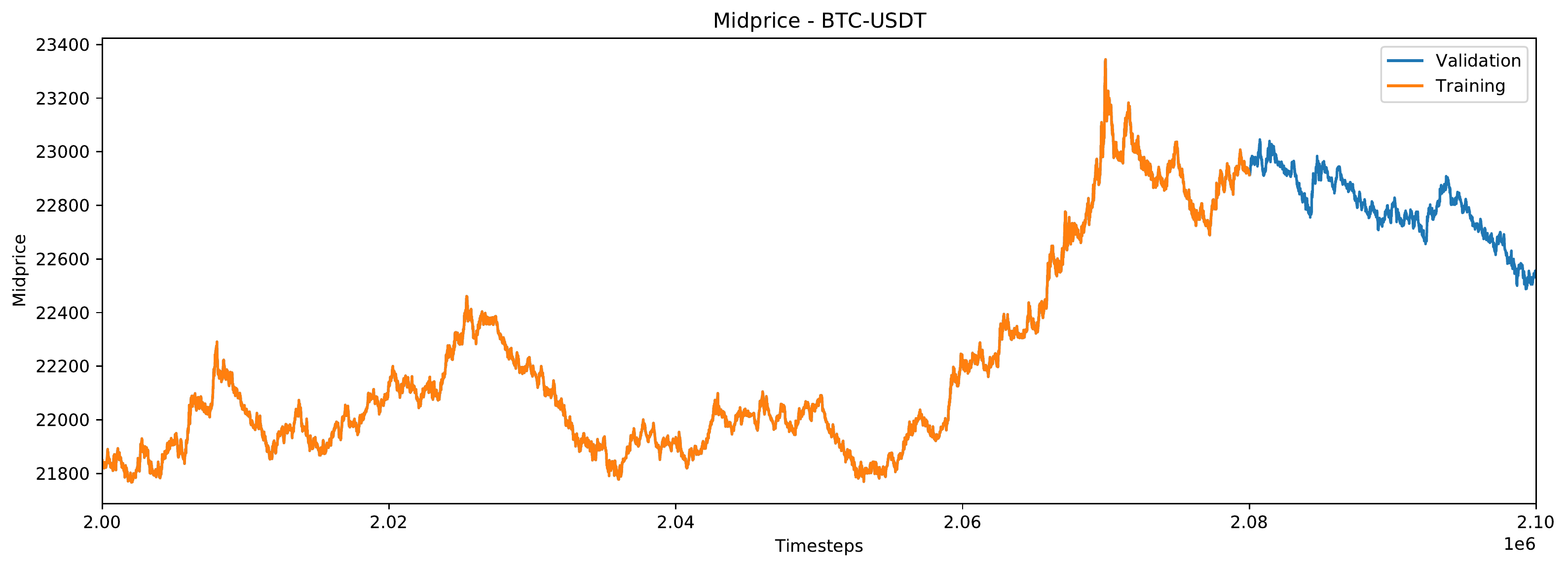}
    \caption{Subset of LOB snapshots used to assess model performance - Midprice.}
    \label{fig3_4_1}
  \end{minipage}
  \hfill
  \begin{minipage}[b]{0.495\hsize}
    \includegraphics[width=\textwidth]{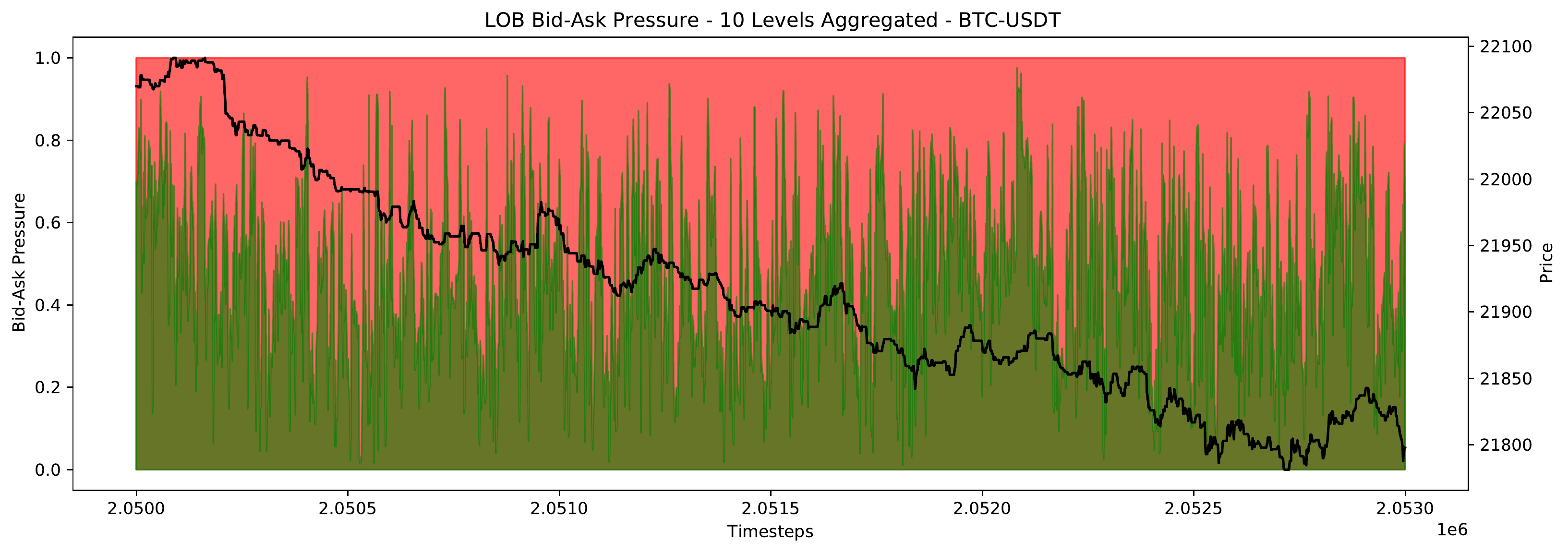}
    \caption{Subset of LOB snapshots used to assess model performance - Bid-Ask pressure.}
    \label{fig3_4_2}
  \end{minipage}
\end{figure}

To compare the performance of different deep learning architectures, we use a subset of 100,000 LOB snapshots, where 80,000 LOB snapshots are used for training and 20,000 for validation (cf. \autoref{fig3_4_1} and \autoref{fig3_4_2}).

The same subset of LOB snapshots is used later for comparing different Transformer architectures. 

\subsection{Setup}

The input data fed into the models consists of 38 features:
\begin{itemize}
    \item 9 bid prices and the 9 associated quantities
    \item 9 ask prices and the 9 associated quantities
    \item historical lagged log-return variable $\log\left(\frac{Midprice_t}{Midprice_{t-\tau}}\right)$ where $\tau$ is the forecast horizon
    \item the weighted midprice
\end{itemize}
The model's output is the log-return for a given forecast horizon $\tau \in \{1,...,30\}$. 

The training uses the adaptive moment estimation Adam with Weight Decay (AdamW)~\cite{loshchilov2017decoupled} optimizer. The AdamW optimizer estimates the first and second moment of the gradient and adapts the learning rate for each input feature. The decoupled weight decay improves the convergence to an optimum.

The  Mean Squared Error (MSE) and Mean Average Error (MAE) loss functions were considered for the LSTM, Transformer, Autoformer, FEDformer, and HFformer models. Models trained with the MSE loss, although more sensitive to outliers than the MAE loss, yielded a higher out-of-sample $R^2$ score. 

\subsection{Evaluation}
Out-of-sample $R^2$ is used to assess the performance of the models for forecasting log-returns at different horizons. $R^2$ has the advantage of being interpretable relative to a baseline score. However, $R^2$ is sensitive to outliers.  Finally, after each training epoch, the model's performance is evaluated using training and validation sets to avoid overfitting. 


\subsection{Results}

\begin{figure}[ht]
\centering 
\includegraphics[width=0.8\linewidth]{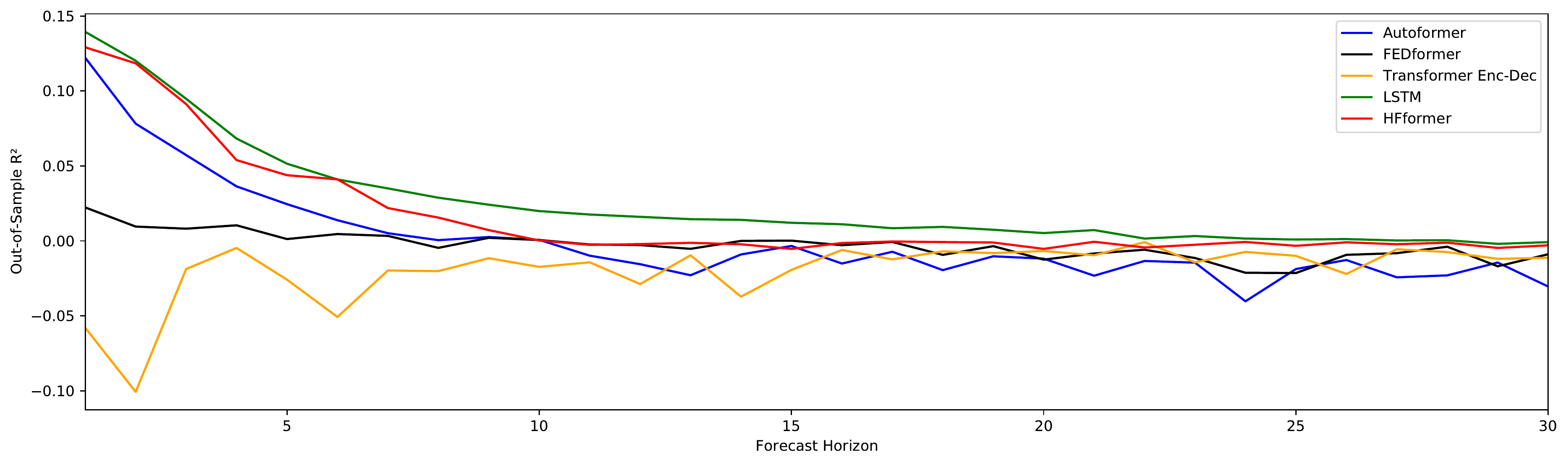} 
\caption{Autoformer vs. FEDformer vs. Transformer vs. LSTM vs. HFformer - Log-returns.}
\label{fig5_2_1}
\end{figure}

From \autoref{fig5_2_1}, we notice that the LSTM outperforms the Transformer model for all forecasting horizons. This could be due to the high noise-to-signal ratio that the LOB contains. The Transformer is more sensitive to noise as it is a more complex neural network which may lead it to overfit the training data even if the training is done using a validation set to reduce the chances of overfitting. We will later see that the Transformer's encoder performance improves with noise-reducing activation functions and larger training sets. 

Among Transformer-like architectures, the Autoformer achieves the best performance on shorter forecasting horizons (cf. \autoref{fig5_2_1}). The Autoformer and FEDformer require an additional input feature which is the date and time timestamp, as both of these models use a time embedding. The frequency of the target time series (e.g., seconds, minutes, and hours) is used as a hyperparameter. 

 The HFformer outperforms the other Transformer-like architectures, however, it underperforms the LSTM (cf. \autoref{fig5_2_1}), however, as the size of the training dataset increases, the HFformer attains better forecasting results than the LSTM during backtesting.

 The hyperparameters are listed in \autoref{tab_4}.

\subsection{Ablation Study}

We conduct an ablation study to determine the impact of different suggested improvements used in the HFformer compared to the original Transformer architecture. 

The following ablations are made:
\begin{itemize}
    \item the spiking activation combined with the PReLU activation function is replaced with a PReLU activation function
    \item the positional encoding is added back
    \item the linear decoder is replaced with the autoregressive Transformer decoder
\end{itemize}

\begin{figure}[ht]
\centering 
\includegraphics[width=0.8\linewidth]{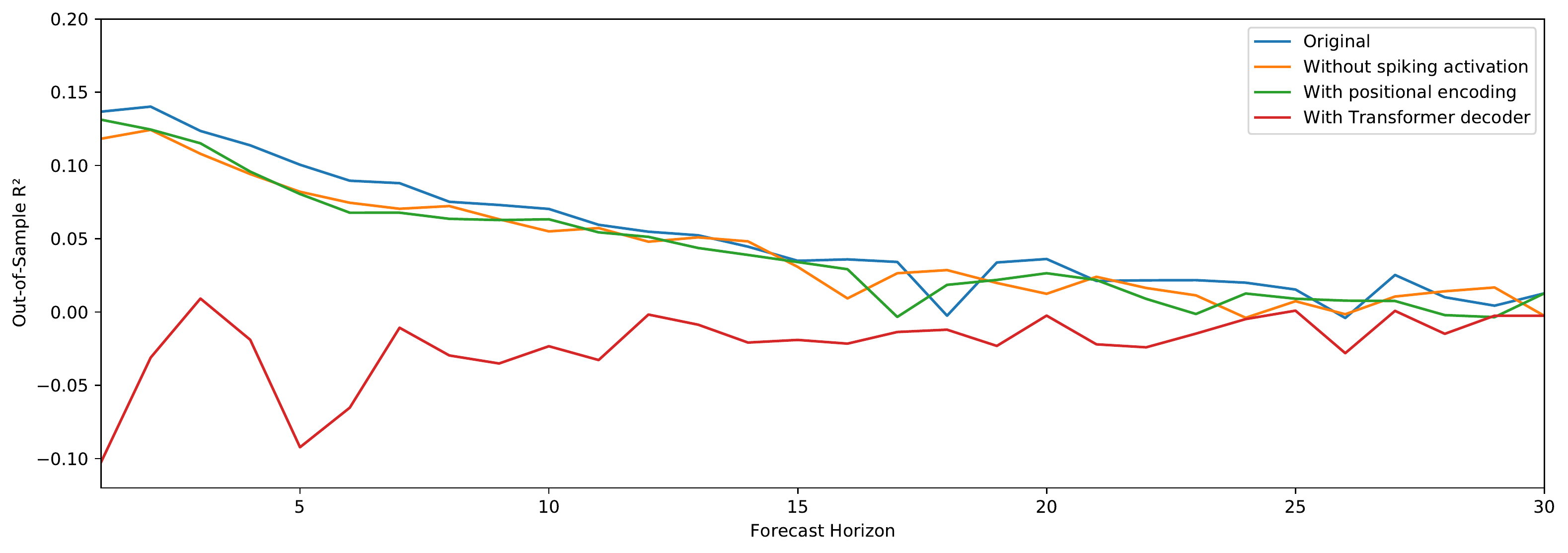} 
\caption{Ablation study for the HFformer - Log-returns.}
\label{fig5_3_1}
\end{figure}

We notice from \autoref{fig5_3_1} that a major improvement in the $R^2$ score is achieved by replacing the Transformer decoder with a linear decoder. Adding a spiking activation and removing the positional encoding result in marginal improvements.

\subsection{Look-back Window: LSTM vs. HFformer}
We experiment with the size of the look-back window to find out which size will yield the best performance for longer forecasting horizons. The LSTM and HFformer models are trained and tested with different look-back window sizes using the same data as previously. 

\begin{figure}[H]
  \centering
  \begin{minipage}[b]{0.495\hsize}
    \includegraphics[width=\textwidth]{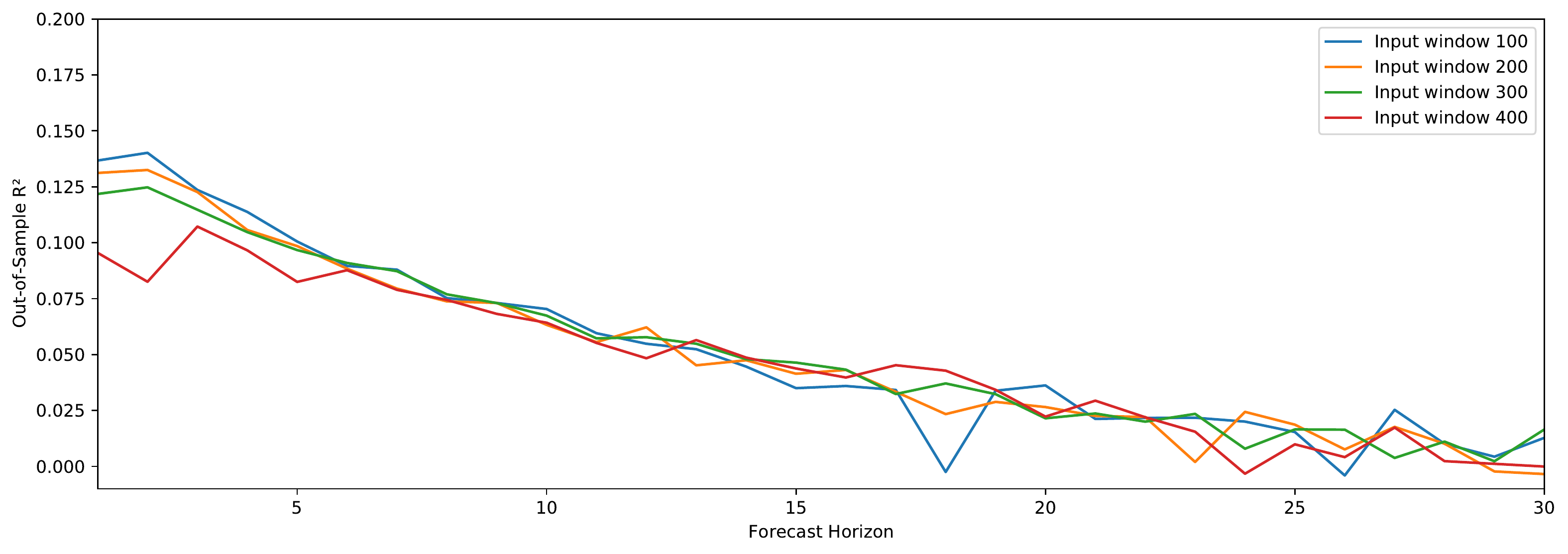}
    \caption{Look-back window size impact on the HFformer - Log returns.}
    \label{fig6_1_1}
  \end{minipage}
  \hfill
  \begin{minipage}[b]{0.495\hsize}
    \includegraphics[width=\textwidth]{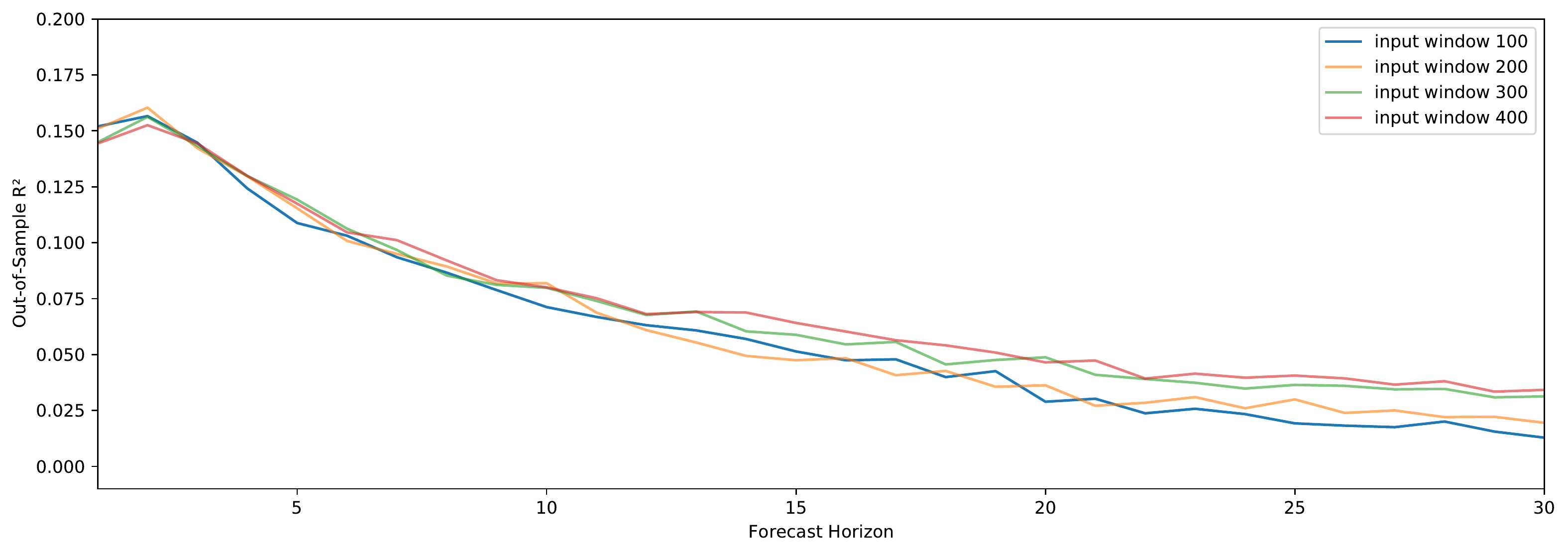}
    \caption{Look-back window size impact on the LSTM - Log-returns.}
    \label{fig6_1_2}
  \end{minipage}
\end{figure}

We notice that increasing the size of the look-back window positively affected the performance of the LSTM (cf. \autoref{fig6_1_2}). However, there was no noticeable change for the HFformer (cf. \autoref{fig6_1_1}).

\subsection{Training and Validation Sets}

The LSTM and HFformer models are trained on sets of the following size:
\begin{itemize}
    \item Training set with 80,000 LOB snapshots and validation set with 20,000 LOB snapshots
    \item Training set with 300,000 LOB snapshots and validation set with 60,000 LOB snapshots
    \item Training set with 600,000 LOB snapshots and validation set with 120,000 LOB snapshots
\end{itemize}

\begin{figure}[H]
\centering 
\includegraphics[width=0.8\linewidth]{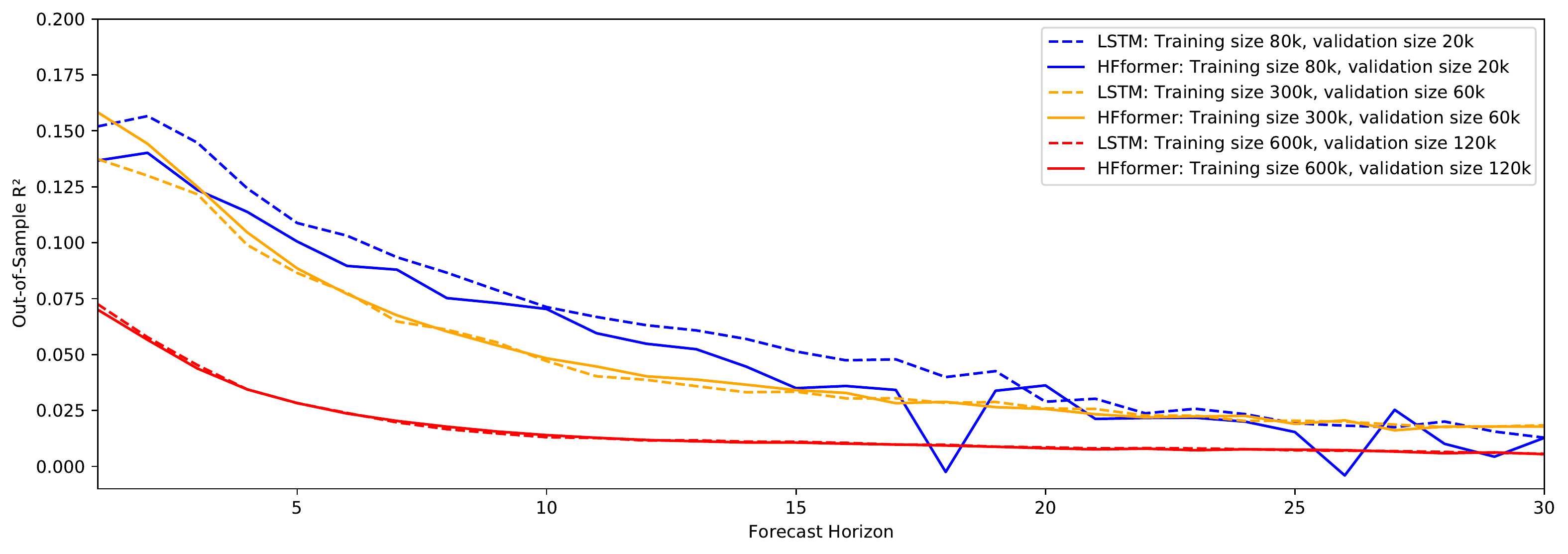} 
\caption{Size of validation and test sets impact on the HFformer and LSTM - Log-returns.}
\label{fig6_2_1}
\end{figure}

From \autoref{fig6_2_1}, we notice that as the training and validation sets increases the absolute performance of both models decreases. However, it is important to consider larger training and validation sets, as during live trading, it can be technically challenging to continuously retrain the model and may lead to overfitting. The LSTM outperforms the HFformer on smaller training and validation datasets. The HFformer's forecasting performance improves relative to the LSTM's as the size of the training and validation sets increases.

\subsection{Classification Performance}

Previous comparisons have been based on $R^2$, which assesses the regressional quality of the model. We now proceed to study the classification performance of the HFformer and LSTM. For the following comparisons, we use the HFformer and LSTM models trained on 600,000 LOB snapshots and validated on 120,000 LOB snapshots. The classification performance is assessed on a test set of size 200,000 LOB snapshots.

We define two classes:
\begin{itemize}
    \item buy, when the predicted log-return is positive
    \item sell, when the predicted log-return is negative
\end{itemize}

The ratios are the true positive rates for each class. 

The first comparison is based on true positives and false positives. The second comparison is a mix of classification and regression. The computed ratios are weighted ratios where the weights are the true values of the log-returns. The motivation behind this comparison is to assess whether both of the models can classify relatively large log-returns correctly. This is important in the case of volatile assets such as Bitcoin since, if the model fails to predict large log-return changes, the trading strategy will result in large drawdowns and limited upsides.

\begin{figure}[H]
  \centering
  \begin{minipage}[b]{0.495\hsize}
    \includegraphics[width=\textwidth]{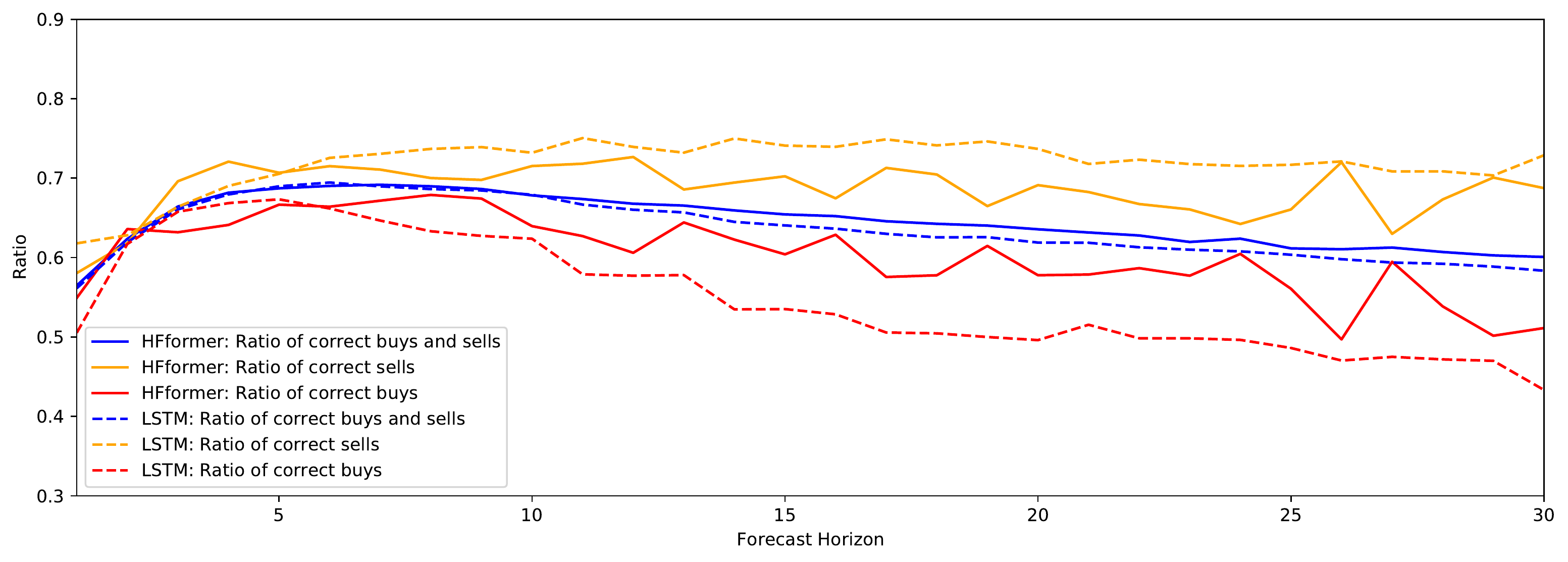}
    \caption{Classification performance of the HFformer and LSTM - Log-returns.}
    \label{fig6_3_1}
  \end{minipage}
  \hfill
  \begin{minipage}[b]{0.495\hsize}
    \includegraphics[width=\textwidth]{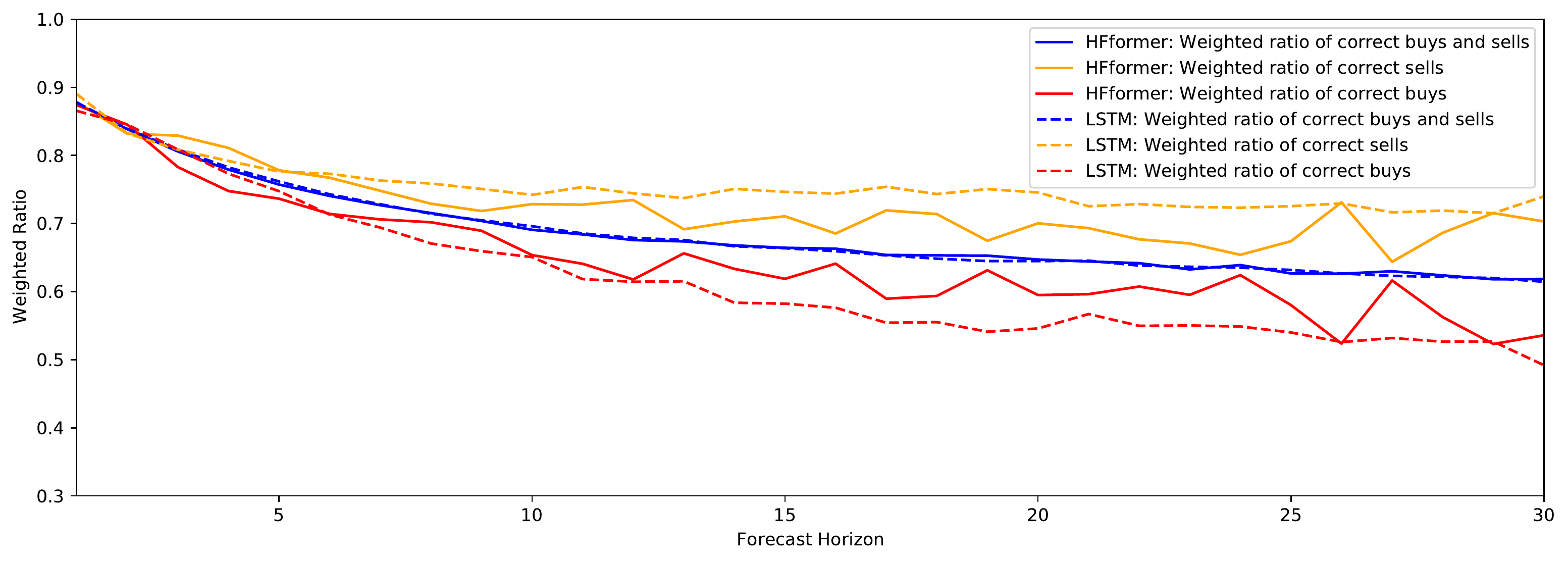}
    \caption{Classification performance with weights of the HFformer and LSTM - Log-returns.}
    \label{fig6_3_2}
  \end{minipage}
\end{figure}

We notice that, on average, the HFformer exhibits a slightly better classification performance than the LSTM (cf. \autoref{fig6_3_1}). For the weighted ratio, the overall performance of the HFformer is similar to the LSTM(cf. \autoref{fig6_3_2}). However, the HFformer has a smaller gap between the correct buys and correct sells ratios. We will later examine the impact of having a model that flags buy and sell opportunities in a more balanced way.

\subsection{Backtesting}
\subsubsection{Strategies}
The HFformer and LSTM models were trained on June 2022 data and are now backtested on BTC-USDT LOB snapshots collected over 2 days (July 21st and 22nd, 2022).


The backtesting is based on the following assumptions:
\begin{itemize}
    \item No trading fee, which is a realistic assumption as there are no spot trading fees on Binance for the BTC-USDT pair
    \item The delay from receiving the snapshots, processing them, and placing a trade is around 200 ms, which is approximately equivalent to 2 ticks for the LOB snapshots that were collected
    \item The orders are market orders
    \item 0.0002\% of the amount bought then sold for long trades (and vice-versa for short trades) is subtracted from the Profit and Loss (PnL) to simulate price slippage
    \item A fixed quantity of 0.1 BTC is traded
\end{itemize}

The following trading strategies are considered:
\begin{itemize}
    \item Strategy 1: the trading horizon is 28 ticks. We use the HFformer and LSTM models trained on the specific horizon to generate a trade signal. If the signal is positive, we open a long position of 0.1 BTC with a delay of 2 ticks and close it after 28 ticks plus 2 ticks of delay. Similarly, if the signal is negative, we open a short position and proceed with a similar method to the long strategy.
    While running strategy 1, when we create a scatter plot of profitable and unprofitable trades based on start and end signals, we notice a relationship between the sign of the signals and the profitability of trades (cf. \autoref{fig6_7_3}).
    \item Strategy 2: the trading horizon is 28 ticks. We use three HFformer and three LSTM models trained on the specific horizon and +/- 2 horizons. Only if all 3 signals are positive, we open a long position of 0.1 BTC with a delay of 2 ticks and close it after 28 ticks plus 2 ticks of delay. Similarly, if all 3 signals are negative, we open a short position and proceed with a similar method to the long strategy.
    \item Strategy 3: similar to Strategy 2, however, instead of using 3 signals to trade, we use 5 signals: signals for horizons 26, 27, 28, 29, and 30. 
\end{itemize}

\subsubsection{Results}
\begin{figure}[H]
\centering 
\includegraphics[width=0.8\linewidth]{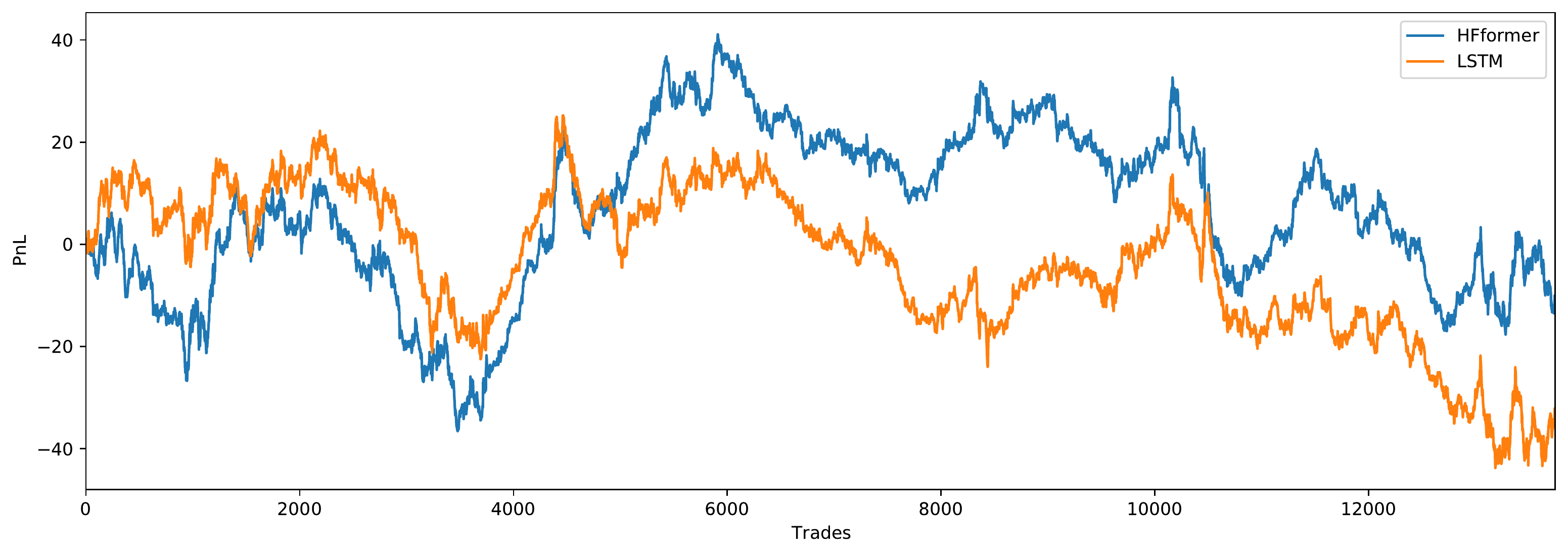} 
\caption{Cumulative PnL - Strategy 1 - HFformer vs. LSTM.}
\label{fig6_7_2}
\end{figure}

\begin{figure}[H]
  \centering
  \begin{minipage}[b]{0.495\hsize}
    \includegraphics[width=\textwidth]{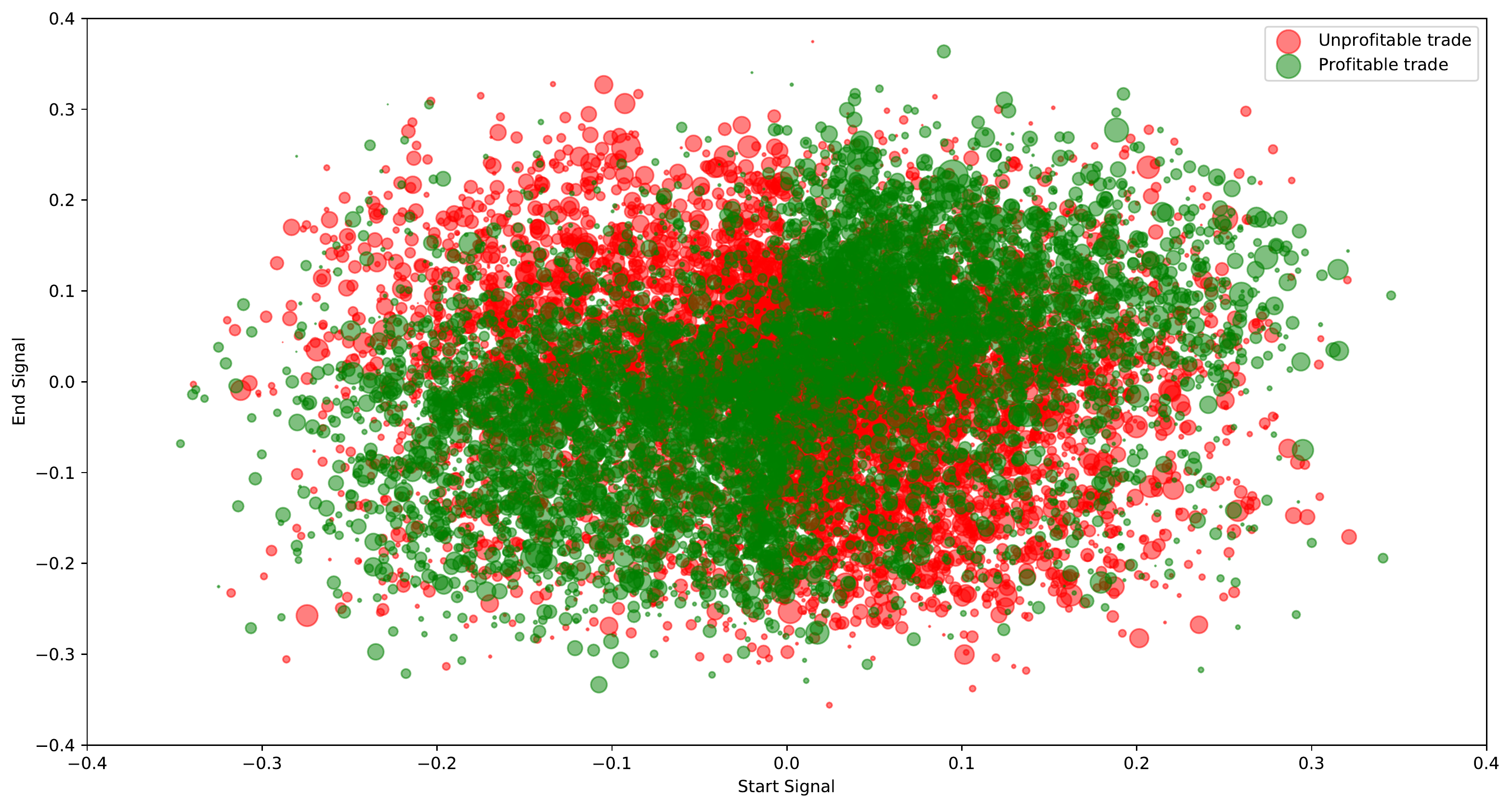}
    \caption{Relationship between PnL and start and end signals outputted by the model - Strategy 1 - HFformer.}
    \label{fig6_7_3}
  \end{minipage}
  \hfill
  \begin{minipage}[b]{0.495\hsize}
    \includegraphics[width=\textwidth]{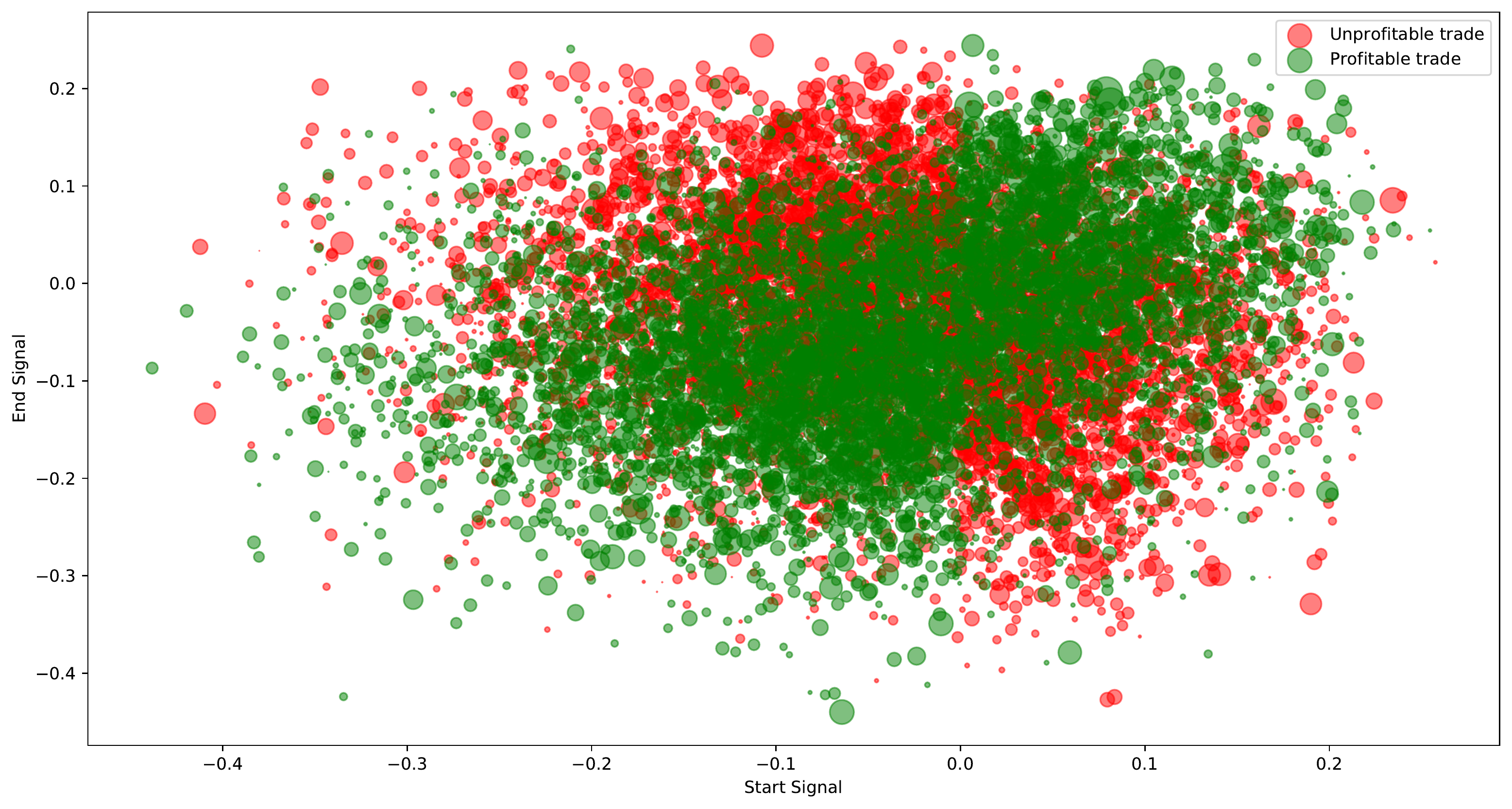}
    \caption{Relationship between PnL and start and end signals outputted by the model - Strategy 1 - LSTM.}
    \label{fig6_7_3_2}
  \end{minipage}
\end{figure}

Strategy 1 results in a final PnL of -10.42 USDT and a total of 13,743 trades for the HFformer and in a final PnL of -32.54 USDT and a total of 13,743 trades for the LSTM. From \autoref{fig6_7_2} and \autoref{fig6_7_3}, we notice that most profitable trades happen when the HFformer model outputs start and end signals of the same sign. One possible explanation is that profitable trades occur when the upward or downward trend predicted by the model lasts longer than the forecasting horizon. Therefore, one possible improvement to this trading strategy is to use multiple forecasting horizons to assess whether to go into a trade. The multiple forecasting horizons can include short forecasting horizons (e.g., 26 and 27), which are shorter than the main forecasting horizon (e.g., 28 in strategy 1) to reduce the chance of falsely flagging a trading opportunity. Additionally, longer forecasting horizons (e.g., 29 and 30) can be used to confirm that the upward or downward trend duration will last beyond the trade delay (e.g., 2 ticks in strategy 1).  This leads us to strategy 2.

Strategy 2 results in a final PnL of 120.04 USDT and a total of 11,928 trades for the HFformer and in a final PnL of 9.67 USDT and a total of 12,932 trades for the LSTM. We notice an improvement in the PnL and a reduction in the total number of trades. The total number of trades decreases as all three models with different forecast horizons need to output a signal of the same sign for a trade to occur. 

\begin{figure}[H]
\centering 
\includegraphics[width=0.8\linewidth]{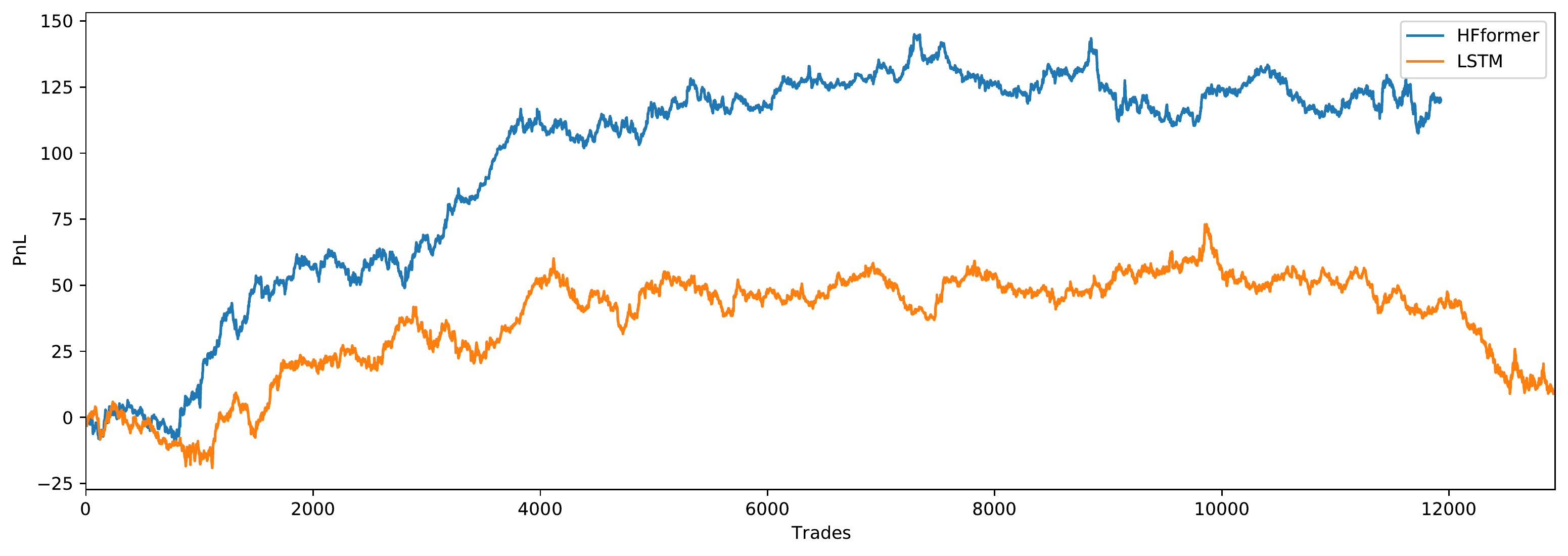} 
\caption{Cumulative PnL - Strategy 2 - HFformer vs. LSTM.}
\label{fig6_7_4}
\end{figure}

From \autoref{table6_7_1}, we see that the correlation between adjacent signals is high but not 1. This allows exploring the option of adding more signals and reducing the number of potentially unprofitable trades by only placing trades when the same signal is exhibited for more models. 

\begin{table}[H]
\begin{center}
\begin{tabular}{lllll}
  \hline
  & \textbf{PnL} & \textbf{Signal - 26} & \textbf{Signal - 28} & \textbf{Signal - 30} \\ 
  \hline
  \textbf{PnL}         &  1        &          &  & \\ 
  \textbf{Signal - 26} & 0.002603  & 1        &  &   \\ 
  \textbf{Signal - 28} & 0.007165  & 0.938949 & 1 &   \\ 
  \textbf{Signal - 30} & -0.002530 & 0.852447 & 0.867870  &  1  \\ 
   \hline
\end{tabular}
\end{center}
\caption{Correlation table for PnL and signals for forecast horizons of 26, 28, and 30 ticks - HFformer.}
\label{table6_7_1}
\end{table}

\begin{table}[H]
\begin{center}
\begin{tabular}{lllll}
  \hline
  & \textbf{PnL} & \textbf{Signal - 26} & \textbf{Signal - 28} & \textbf{Signal - 30} \\ 
  \hline
  \textbf{PnL}         &  1        &          &  & \\ 
  \textbf{Signal - 26} & 0.000514  & 1        &  &   \\ 
  \textbf{Signal - 28} & 0.005031  & 0.968979 & 1 &   \\ 
  \textbf{Signal - 30} & 0.001102 & 0.972374 & 0.970612  &  1  \\ 
   \hline
\end{tabular}
\end{center}
\caption{Correlation table for PnL and signals for forecast horizons of 26, 28, and 30 ticks - LSTM.}
\label{table6_7_1_2}
\end{table}

Strategy 3 results in a final PnL of 138.78 USDT and a total of 10,976 trades for the HFformer. By adding two signals, the cumulative PnL has improved by 15.61\% compared to strategy 2. As expected, the number of trades decreased by 7.98\% compared to strategy 2 and 20.13\% compared to strategy 1. Strategy 3 results in a final PnL of 85.71 USDT and a total of 12,226 trades for the LSTM.  By adding two signals, the cumulative PnL has improved nine-fold. As expected, the number of trades decreased by 5.46\% compared to strategy 2 and 11.04\% compared to strategy 1.

\begin{figure}[H]
\centering 
\includegraphics[width=0.8\linewidth]{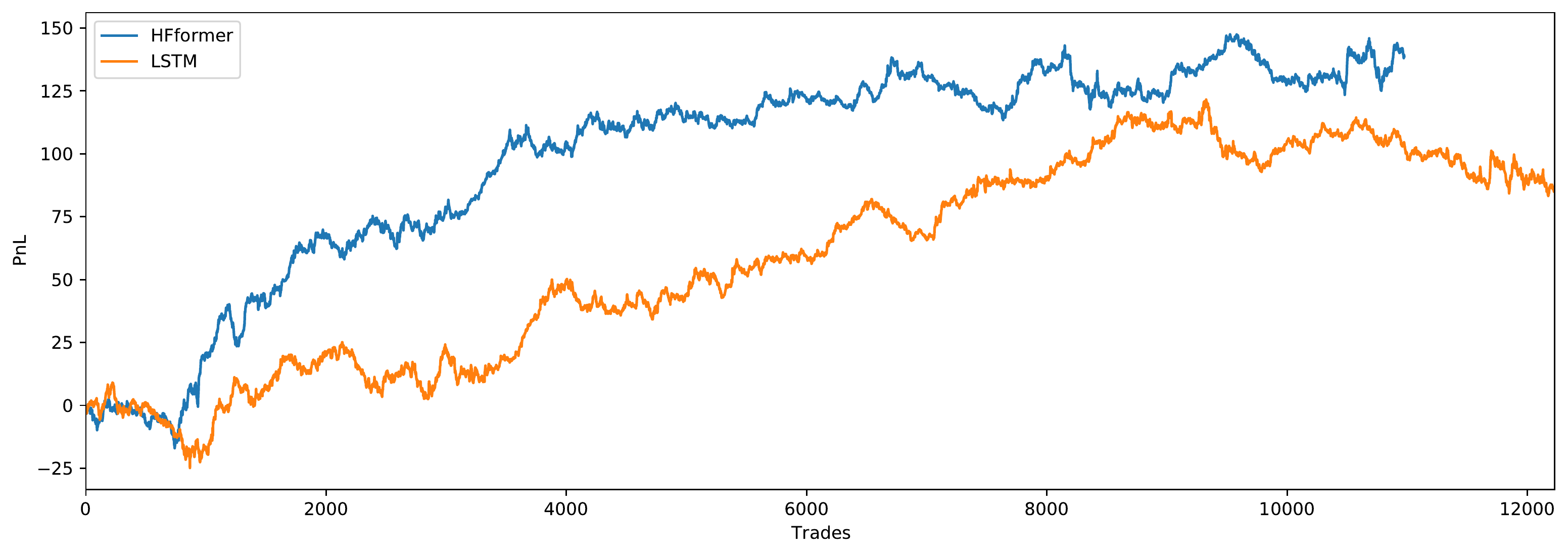} 
\caption{Cumulative PnL - Strategy 3 - HFformer vs. LSTM.}
\label{fig6_7_5}
\end{figure}

Comparing \autoref{table6_2_1} and \autoref{table6_2_1_2}, we notice that the signals for different forecasting horizons of the HFformer are less correlated than the signals generated by the LSTM. Therefore when HFformer signals are combined, we get fewer trades, but these trades exhibit stronger buy or sell signals. As a result, the HFformer signals, on average, result in a higher PnL than the LSTM signals (cf. \autoref{fig6_7_4} and \autoref{fig6_7_5}).

\begin{table}[H]
\begin{center}
\begin{tabular}{lllllll}
  \hline
              & \textbf{PnL}        & \textbf{Signal - 26}  & \textbf{Signal - 27} & \textbf{Signal - 28} & \textbf{Signal - 29} & \textbf{Signal - 30}\\ 
              \hline
  \textbf{PnL}         &            1 &           &           &           &          &   \\ 
  \textbf{Signal - 26} & 	0.009710 &         1 &           &           &          &   \\ 
  \textbf{Signal - 27} &  	0.013158 &  0.943009 & 1         &           &          &   \\ 
  \textbf{Signal - 28} &     0.004809 &  0.860272 & 0.877412  &  1        &          &   \\ 
  \textbf{Signal - 29} &  	0.014907 &  0.923525 & 0.932065  &  0.858861 &         1&   \\ 
  \textbf{Signal - 30} &     0.010281 &  0.914158 & 0.919749  &  0.861871 &  0.912567& 1 \\ 
   \hline
\end{tabular}
\end{center}
\caption{Correlation table for PnL and signals for forecast horizons of 26, 27, 28, 29, and 30 ticks - HFformer.}
\label{table6_2_1}
\end{table}

\begin{table}[H]
\begin{center}
\begin{tabular}{lllllll}
  \hline
              & \textbf{PnL}        & \textbf{Signal - 26}  & \textbf{Signal - 27} & \textbf{Signal - 28} & \textbf{Signal - 29} & \textbf{Signal - 30}\\ 
              \hline
  \textbf{PnL}         &            1 &           &           &           &          &   \\ 
  \textbf{Signal - 26} & 	0.007214 &         1 &           &           &          &   \\ 
  \textbf{Signal - 27} &  	0.011843 &  0.971106 & 1         &           &          &   \\ 
  \textbf{Signal - 28} &     0.008985 &  0.974401 & 0.972183  &  1        &          &   \\ 
  \textbf{Signal - 29} &  	0.007163 &  0.957233 & 0.956653  &  0.953691 &         1&   \\ 
  \textbf{Signal - 30} &     0.010961 &  0.965258 & 0.968503  &  0.964354 &  0.955223& 1 \\ 
   \hline
\end{tabular}
\end{center}
\caption{Correlation table for PnL and signals for forecast horizons of 26, 27, 28, 29, and 30 ticks - LSTM.}
\label{table6_2_1_2}
\end{table}

\begin{figure}[H]
\centering 
\includegraphics[width=0.8\linewidth]{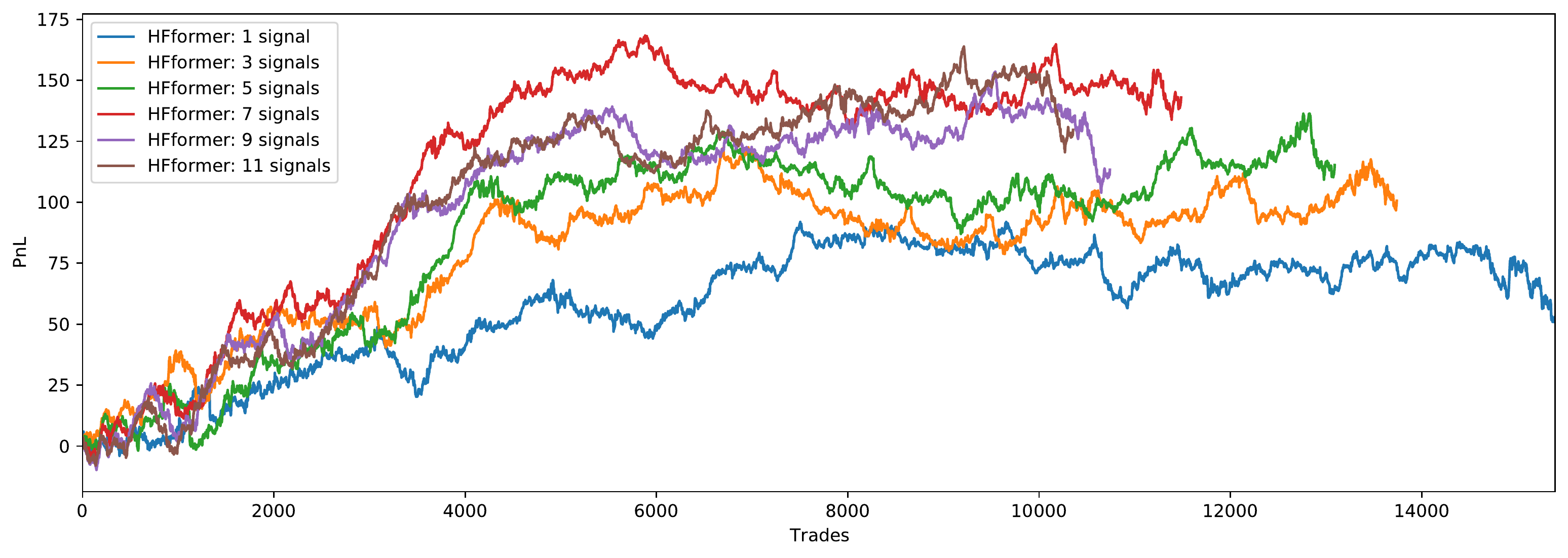} 
\caption{Cumulative PnL - Main horizon 25 - HFformer.}
\label{fig6_7_6}
\end{figure}

We assess HFformer's performance using 1 to 11 trading signals (cf. \autoref{fig6_7_6}). The predictions are made for horizon 25 and signals are added progressively by an increment of 1 from each side. We notice an improvement in the cumulative PnL when adding signals. However, after 7 signals the cumulative PnL stagnates. Therefore, for the following experiments we will be using 7 signals.

\subsubsection{More on Trading Signal Aggregation}

\begin{figure}[H]
\centering 
\includegraphics[width=0.8\linewidth]{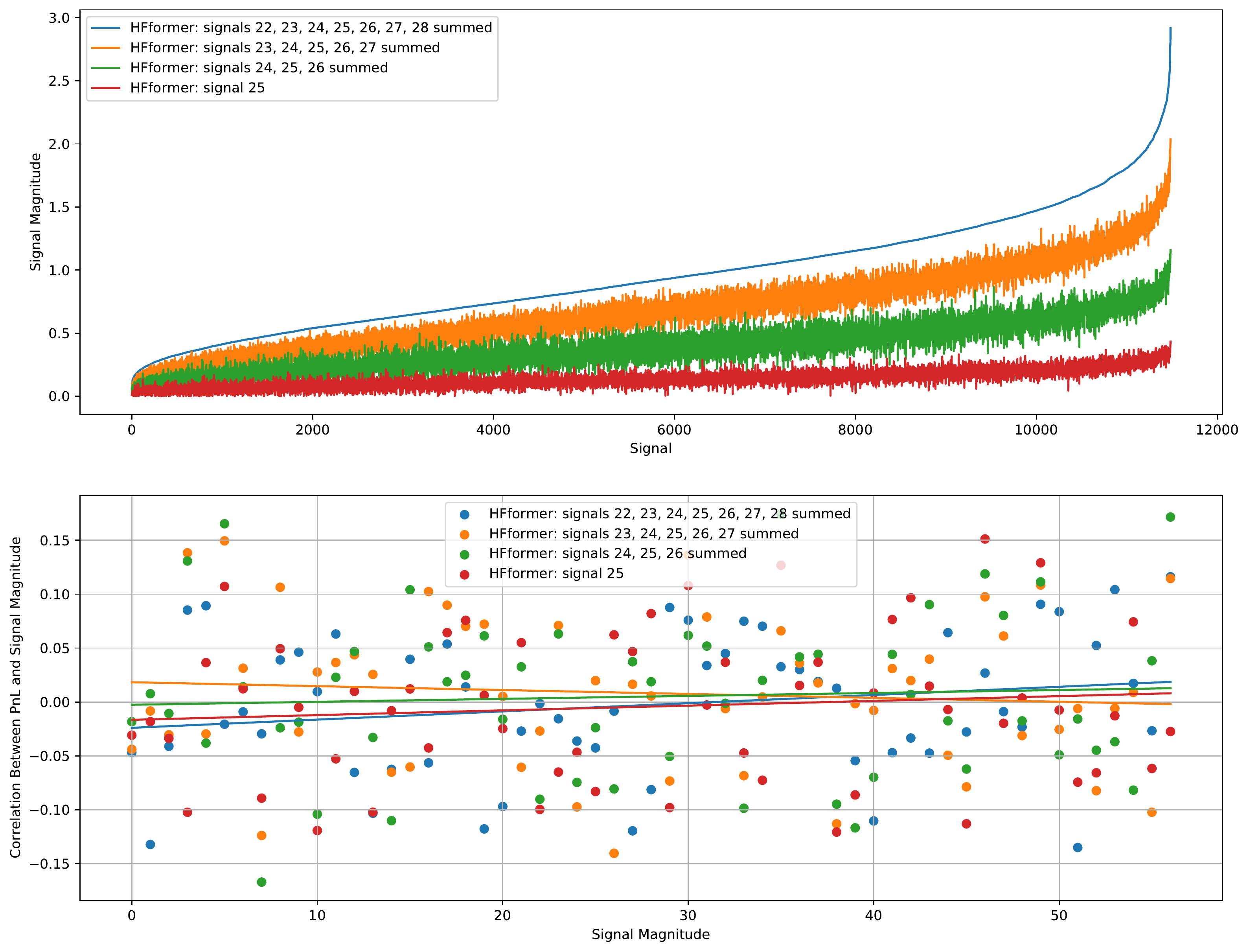} 
\caption{Trade signal magnitude (top) and correlation between PnL and trade signal magnitude (bottom) - Trade sizing - Main forecasting horizon 25 with 1, 3, 5 and 7 signals - HFformer.}
\label{fig6_7_9}
\end{figure}

From \autoref{fig6_7_9}, we notice that as we aggregate more signals through summation and then take the absolute value of this sum, the correlation between the PnL and the aggregated signals' magnitude increases (except for the case with five aggregated signals plotted in orange). To perform this comparison, we compute the magnitudes of the aggregated signals and then sort the signals in ascending order. We then compute the correlation between the signal magnitudes and the PnL by batches of 200 observations. Based on these observations, we can improve the trading strategies presented above by using aggregated sums of signals instead of only focusing on the sign of the signals. Additionally, we can disregard trading signals below a certain minimum threshold.  

\subsubsection{Trade Sizing}

\begin{figure}[H]
\centering 
\includegraphics[width=0.8\linewidth]{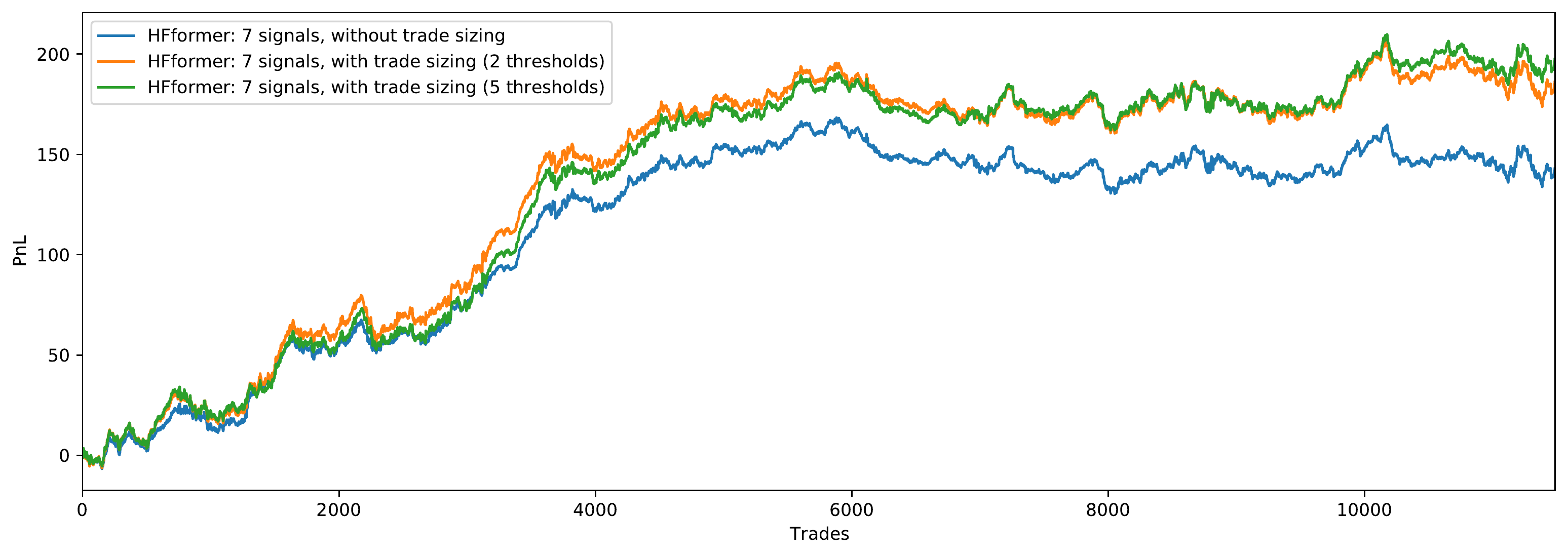} 
\caption{Cumulative PnL - Trade sizing - Main forecasting horizon 25, 7 signals, and 2 and 5 trade sizing thresholds - HFformer.}
\label{fig6_7_7}
\end{figure}

Finally, we assess HFformer's performance using trade sizing (cf. \autoref{fig6_7_7}). The trade sizing is done depending on the absolute value of the sum of the trade signals. We use two sets of thresholds:
\begin{itemize}
    \item 2 thresholds: above the highest threshold 0.15 BTC is traded, 0.1 BTC for the medium threshold, and 0.05 BTC otherwise
    \item 5 thresholds: above the highest threshold, 0.15 BTC is traded, 0.125 BTC for the second highest threshold, 0.1 BTC for the third highest threshold, 0.075 for the fourth highest threshold, 0.05 for the fifth highest threshold, and 0.025 BTC otherwise
\end{itemize}
The thresholds above were chosen by observing the distribution of outputted signals using training data.

For this experiment, we use 7 signals with a main forecasting horizon of 25 ticks. We notice an increase in the cumulative PnL when using trade sizing. The standard deviation of the cumulative PnL increases from 0.49 USDT without trade sizing to 0.58 USDT with trade sizing using 2 thresholds. However, when the number of thresholds for trade sizing is increased to 5, the PnL does not increase significantly, but the standard deviation of the PnL drops to 0.41 USDT. A lower standard deviation of the PnL will yield a better Sharpe ratio long-term and reduce the volatility of the high-frequency strategy.

\begin{figure}[H]
\centering 
\includegraphics[width=0.8\linewidth]{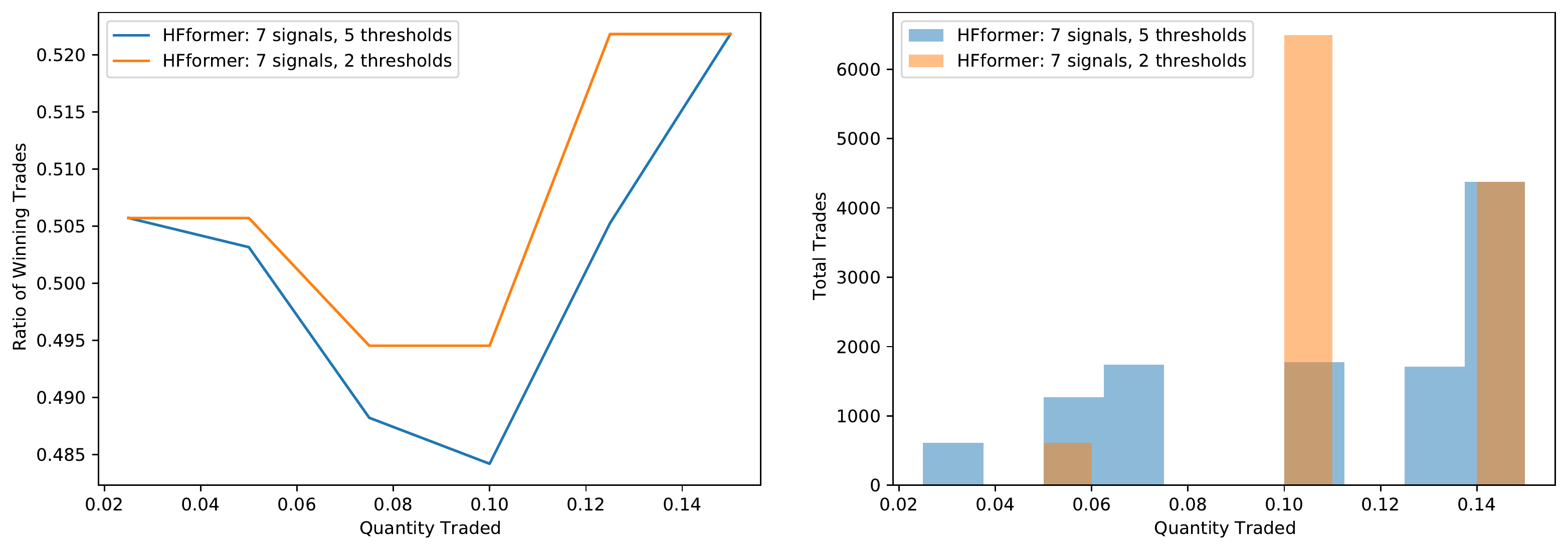} 
\caption{Wining trades ratio - Trade sizing - Main forecasting horizon 25, 7 signals, and 2 and 5 trade sizing thresholds - HFformer.}
\label{fig6_7_8}
\end{figure}

Moreover, from \autoref{fig6_7_8}, we notice that the ratio of trades with a positive PnL over the total number of trades (i.e., the ratio of winning trades) increases as the magnitude of the trading signal increases in line with previous observations. Therefore, one proposed improvement to the trading strategy is to increase the quantity traded proportionally to the strength of the trading signal. 

\begin{figure}[H]
\centering 
\includegraphics[width=0.8\linewidth]{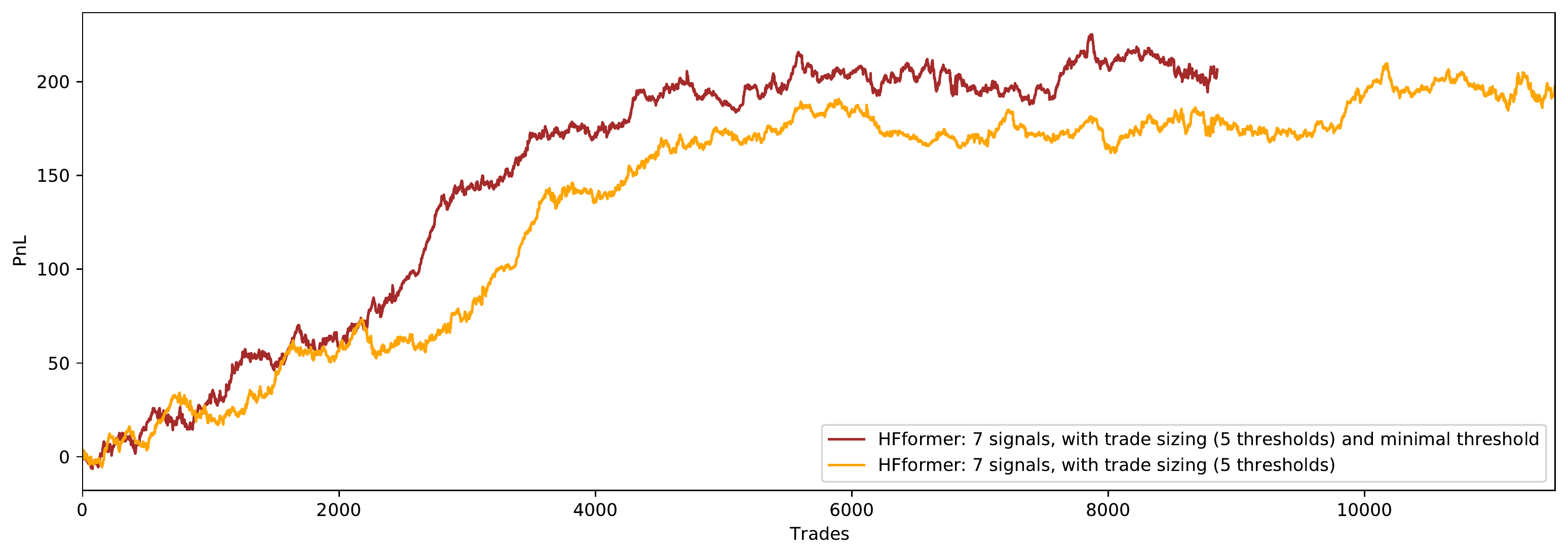} 
\caption{Cumulative PnL - Trade sizing with minimal threshold - Main forecasting horizon 25, 7 signals, and 5 trade sizing thresholds - HFformer.}
\label{fig6_7_10}
\end{figure}

Finally, we complement the existing trading strategy from \autoref{fig6_7_7} with a minimal threshold requirement for a trade to occur. The minimal threshold is set using training data observations. As previously, we compute the sum of trading signals from the models and take the absolute value. The minimal trading threshold is proportional to the number of signals. From \autoref{fig6_7_10} we notice an improvement in the cumulative PnL by 4.53\% and a decrease in the number of trades from 11,485 to 8,851.

\section{Legal and Ethical Considerations}
The strategies and models presented in this paper are purely for scientific research and curiosity. Algorithmic trading may result in capital loss and requires continuous risk monitoring. The strategies should first be paper traded, and various risks should be assessed (e.g., price, liquidity, compliance, and regulatory risks) before being deployed in live trading. Simplifications to the trading environment have been made during backtesting for this paper. Therefore, results may vary significantly if the models are deployed live. Moreover, machine learning methods are evaluated based on classification and regression metrics without rigorous mathematical proofs, unlike statistical methods. As a result, machine learning methods' performance can significantly vary depending on the input data and could lead to unexpected model outputs.  

Algorithmic trading is regulated by financial authorities (e.g., Financial Industry Regulatory Authority and Security Exchange Commission for US securities). Regulatory risks need to be considered and monitored when trading in financial markets. The trading strategies need to be reviewed and monitored to be in accordance with continuously changing market regulations.

\section{Limitations and Future Lines of Research}

The HFformer and LSTM results were achieved using a simplified backtesting environment, which did not account for factors including:
\begin{itemize}
    \item the impact of the executed order on the market
    \item the actual execution price; this is set to the weighted midprice in our model
    \item the available BTC quantities; the executed quantity is 0.1 BTC 
    \item any connection issues; the data was fed into the model at the same rate
    \item trading fees; the trading fee was set to zero, which is a realistic assumption for the BTC-USDT pair on Binance. However, trading on traditional exchanges usually requires paying a transaction fee
    \item a longer backtesting period; the backtesting period is 2 days 
\end{itemize}

The following future lines of research can be undertaken:
\begin{itemize}
    \item Improve the LOB snapshot pre-processing pipeline to reduce the noisiness of the data with automated feature selection by using autoencoders \cite{han2018autoencoder}
    \item Perform a more extensive performance assessment of the HFformer on large forecast horizons and use altcoin trading pairs such as ETH-USDT
    \item Implement the HFformer with other types of Attention modules such as auto-correlation Attention \cite{xu2021autoformer}
    \item Implement a more realistic backtesting environment that accounts for the impact of the placed order and emulates the activity of other participants in the market to assess the performance of the HFformer
    \item  Test the HFformer's forecasting performance on traditional financial data by using LOB data from other papers focused on LOB deep learning models for midprice forecasting of traditional assets\cite{ kolm2021deep}
\end{itemize}
\section{Conclusion}

This paper studied the LSTM, Transformer, Autoformer, FEDformer, TFT, and HFformer deep learning architectures for high-frequency FTS forecasting. 

Through experimentation, we combined various components of multiple Transformer architectures to form the HFformer, a Transformer-like architecture adapted for HFT. When testing the LSTM and HFformer models, we achieved a higher $R^2$ score than the other deep learning architectures for log-returns forecasting from 1 to 30 ticks ahead. Moreover, the LSTM and HFformer models achieved similar performance for classification tasks. Finally, the LSTM and HFformer models were backtested on different trading strategies involving 1, 3, and 5 trade signals. As a result, it was found that using more than 1 trade signal decreases the number of trades and increases the cumulative PnL of a long-short trading strategy. 

The HFformer, which uses the multi-head Attention mechanism, generates long and short trade signals that result in a more balanced trading strategy than the LSTM. Moreover, the trade signals generated by the HFformer for different forecasting horizons are less correlated than the LSTM signals. As a result, a trading strategy that combines multiple trade signals around a given forecasting horizon was proposed. This trading strategy reduces the false positive trade signals by only engaging in trades when all signals are of the same sign, which results in higher cumulative PnLs and fewer trading fees. 

Finally, the proposed trading strategies were complemented with trade sizing to improve the cumulative PnL. By using different trading quantities based on the strength of the trading signal, the cumulative PnL increased, and in some cases, the volatility of the trading strategy decreased. Additionally, ignoring trade signals below a predefined minimal threshold reduced the total number of trades and positively contributed to the average trade PnL. 

Although these improvements and strategies have been backtested on a large amount of BTC-USDT LOB data collected over 2 days and a month after the training and validation data, these methods may yield different results when trading another cryptocurrency pair or financial asset. Additionally, machine learning methods may sometimes be less generalizable than traditional statistical methods as the machine learning methods are data-driven. 

\bibliographystyle{vancouver}
\bibliography{progressreport.bib}
The following Github repositories were used to build the deep learning models:
\begin{itemize}
    \item LSTM: \href{https://github.com/lkulowski/LSTM\_encoder\_decoder}{https://github.com/lkulowski/LSTM\_encoder\_decoder}
    \item Transformer: \href{https://github.com/AIStream-Peelout/flow-forecast}{https://github.com/AIStream-Peelout/flow-forecast}
    \item Autoformer and FEDformer: \href{https://github.com/cure-lab/DLinear}{https://github.com/cure-lab/DLinear}
    \item Spiking activation: \href{https://github.com/nengo/pytorch-spiking}{https://github.com/nengo/pytorch-spiking}
\end{itemize}

\section{Appendix}\label{appendix}

The training of the deep learning models was done on Google Colab\footnote{The Google Colab platform can be found at \href{https://colab.research.google.com/notebooks/}{https://colab.research.google.com/notebooks/}.} using an Nvidia P100 GPU. The training LOB data for the USDT-BTC trading pair collected from Binance's API can be found below:
\begin{itemize}
    \item raw LOB data: https://drive.google.com/drive/folders/1GXXiVyXXCXenNsGWRAmMb Xd1Xvf-lZq5?usp=sharing
    \item processed LOB data: https://drive.google.com/drive/folders/1GVJ050lVeS6vFjZ9CER YkL5W5YhKOLHw?usp=sharing
\end{itemize}

The hyperparameters that were used to train the LSTM, Transformer, Autoformer, FEDformer, and HFformer models were found through grid search and are listed below:

\begin{table}
  \centering
  \begin{tabular}{p{0.14\linewidth} | p{0.2\linewidth} | p{0.16\linewidth} | p{0.2\linewidth} | p{0.1\linewidth} | p{0.1\linewidth}}
    \toprule
    \textbf{Model} & \textbf{Layers} & \textbf{Activation Function} & \textbf{Optimizer} & \textbf{Loss Function} & \textbf{Batch Size} \\
    \midrule
    LSTM & 5 layers size 16 & PReLU & AdamW with 0.001 learning rate & MSE & 64 \\
    \addlinespace
    Transformer & 1 encoder and 1 decoder both size 128 with 8 Attention heads & PReLU & AdamW with 0.0001 learning rate & MSE & 64 \\
    \addlinespace
    Autoformer & 2 encoders and 2 decoders both of size 128 with 8 Attention heads, label length 50, and factor 3 & GELU & AdamW with 0.0002 learning rate & MSE & 256 \\
    \addlinespace
    FEDformer & 2 encoders and 2 decoders both of size 128 with 8 Attention heads, label length 50, and Legendre base & GELU & AdamW with 0.0005 learning rate & MSE & 256 \\
    \addlinespace
    HFformer & 1 Transformer encoder and 1 linear decoder both size 64 with 6 Attention heads & Spiking with PReLU & AdamW with 0.04 learning rate & MSE & 256 \\
    \bottomrule
  \end{tabular}
  \caption{LSTM, Transformer, Autoformer, FEDformer, and HFformer hyperparameters.}
  \label{tab_4}
\end{table}


    



\end{document}

%% file: includes.tex
\usepackage[numbers]{natbib}
\usepackage{tabularx,longtable,multirow,subfigure,caption}
\usepackage{fancyhdr} 
\usepackage{url} 
\usepackage[english]{babel}
\usepackage{amsmath}
\usepackage{graphicx}
\usepackage{dsfont}
\usepackage{epstopdf} 
\usepackage{backref} 
\usepackage{array}
\usepackage{latexsym}
\usepackage[pdftex,pagebackref,hypertexnames=false,colorlinks]{hyperref} 

\hypersetup{pdftitle={},
  pdfsubject={}, 
  pdfauthor={},
  pdfkeywords={}, 
  pdfstartview=FitH,
  pdfpagemode={UseOutlines},
  bookmarksnumbered=true, bookmarksopen=true, colorlinks,
    citecolor=black,%
    filecolor=black,%
    linkcolor=black,%
    urlcolor=black}

\usepackage[all]{hypcap}

\def\@makechapterhead#1{%
  \vspace*{10\p@}%
  {\parindent \z@ \raggedright \sffamily
    \interlinepenalty\@M
    \Huge\bfseries \thechapter \space\space #1\par\nobreak
    \vskip 30\p@
  }}

\def\@makeschapterhead#1{%
  \vspace*{10\p@}%
  {\parindent \z@ \raggedright
    \sffamily
    \interlinepenalty\@M
    \Huge \bfseries  #1\par\nobreak
    \vskip 30\p@
  }}

\allowdisplaybreaks

%% file: notation.tex
\newcommand{\R}[0]{\mathds{R}} 
\renewcommand{\vec}[1]{{\boldsymbol{{#1}}}} 